\documentclass[useAMS,usenatbib]{mn2e}
\usepackage[british]{babel}

\usepackage{graphicx}
\usepackage{amsmath}
\usepackage{mathabx}

\title[CGM at $z \sim 3$]{A high baryon fraction in massive haloes at $z \sim 3$}
\author[Pezzulli and Cantalupo]{Gabriele Pezzulli$^{1}$\thanks{E-mail:gabriele.pezzulli@phys.ethz.ch} and Sebastiano Cantalupo$^{1}$\\
$^{1}$ Department of Physics, ETH Zurich, Wolfgang-Pauli-Strasse 27, 8093 Zurich, Switzerland \\
}

\begin{document}

\date{Accepted 2019 March 22. Received 2019 March 18; in original form 2018 October 1}

\pagerange{\pageref{firstpage}--\pageref{lastpage}} \pubyear{2019}

\maketitle

\label{firstpage}

\begin{abstract}
\noindent
We investigate the baryon content of the circumgalactic medium (CGM) within the virial radius of $M_\textrm{h} \sim 10^{12} \; \textrm{M}_\odot$ haloes at $z \sim 3$, by modelling the surface brightness profile of the giant Ly$\alpha$ nebulae recently discovered by MUSE around bright QSOs at this redshift. We initially assume fluorescent emission from cold photo-ionized gas confined by the pressure of a hot halo. Acceptable CGM baryon fractions (equal to or smaller than the cosmological value) require that the cold gas occupies $\lesssim 1\%$ of the volume, but is about as massive as the hot gas. CGM baryon fractions as low as 30\% of the cosmic value, as predicted by some strongly ejective feedback models at this redshift, are not easy to reconcile with observations, under our assumptions,  unless \emph{both} the QSO-hosting haloes at $z\sim3$ are more massive than recent BOSS estimates based on clustering \emph{and} the photo-ionized gas is colder than expected in a standard QSO ionizing radiation field. We also consider the option that the emission is dominated by photons scattered from the QSO broad-line region. In this scenario, a very stringent lower limit to the baryon fraction can be obtained under the extreme assumption of optically thin scattering. We infer in this case a baryon fraction of at least 70\% of the cosmic value, for fiducial parameters. Lower values require halo masses or gas temperatures different than expected, or that some mechanism keeps the cold gas systematically overpressured with respect to the ambient medium.

\end{abstract}


\section{Introduction}\label{sec::Introduction}

Modern galaxy formation models rely on means to keep outside of the star-forming body of galaxies most of the baryons theoretically associated with their haloes. Different models achieve this result in different ways. In particular, the so-called \emph{missing} baryons can be stored either relatively close to galaxies (in the circumgalactic medium, or CGM) or very far away from them (in the intergalactic medium, or IGM).\footnote{Here and afterwards, we \emph{define} the boundary between the CGM and the IGM to be the virial radius.}
As a consequence, observational estimates of the (baryonic) mass of the CGM, as a function of halo mass and redshift, are of great importance to further constrain existing models and guide advancement of galaxy formation theory.

Shock heating as a consequence of gravitational collapse was first identified as a possible means to bring most of the baryons falling into the potential wells of forming galaxies to the virial temperature, leading to the formation of a low-density, hot and massive CGM  (\citealt{ReesOstriker1977}; \citealt{WhiteRees1978}). Whenever the dynamical time is long compared to the cooling time, however, the process is largely compensated by radiative cooling, which would lead to the collapse of most of the baryons towards the centre and to excessive star formation compared to observations (e.g.\ \citealt{WhiteFrenk1991}). In some conditions, a virial shock does not even develop, or it is unstable, and filaments of cold gas can penetrate deep into the halo, possibly fuelling substantial star formation at the centre (e.g.\ \citealt{BirnboimDekel2003}; \citealt{Binney2004}; \citealt{Keres+05}). The conditions for efficient radiative cooling are far from trivial, as they crucially depend on metallicity (e.g.\ \citealt{SutherlandDopita1993}), on the intensity and shape of a meta-galactic ionizing background (e.g.\ \citealt{Efstathiou1992}; \citealt{Wiersma+09}), on the time-scales needed to ensure collisional or photo-ionization equilibrium (\citealt{Gnat2017}) and on whether local ionizing sources are consistently taken into account (\citealt{Cantalupo2010}). Some of these effects have been shown to be important, especially at low masses (e.g. \citealt{Kannan+16}), while, at least at high masses, it is most of the times assumed that catastrophic cooling must be compensated by fresh injection of energy, which either provides additional heating to the CGM, or mechanically removes a substantial fraction of the CGM from galaxy haloes. The second option can be referred to as \emph{ejective feedback}.

The most commonly advocated energy sources are supernovae (SNe). First proposed by \cite{DekelSilk1986}, SN-driven \emph{galactic winds} can, if certain conditions are met, eject a large fraction of the baryons out of the potential wells of galaxies and thus solve the overcooling problem. For this reason, they have now become a key ingredient of most semi-analytical galaxy formation models (e.g.\ \citealt{Henriques+13} and references therein), as well as hydrodynamical cosmological simulations (e.g.\ \citealt{SpringelHernquist2003}; \citealt{OppenheimerDave2008}; \citealt{DallaVecchiaSchaye2012}). Expulsion of baryons due to SN feedback is expected to be particularly effective for low-mass galaxies, due to their low escape speed. Crucially, however, the process will impact the baryon content of high-mass haloes as well. Baryons ejected from the shallow potential wells of dwarf galaxies at very high redshift can reach considerably large distances. As a consequence, more massive haloes, formed later from the merging of the low-mass ones, would find themselves relatively deficient in baryons already at virialization. The baryon fraction in these haloes would then slowly increase towards the cosmological value, as the gas ejected at early times is gradually accreted back (`re-incorporated'). With some dependence on implementation details, this scenario provides predictions for the baryon fraction within the virial radius as a function of halo mass and redshift (e.g.\ \citealt{Dave2009}; \citealt{Muratov+15}; \citealt{Liang+16};  \citealt{Mitchell+18}), thus indicating a way to test or refine the paradigm observationally. 

There are reasons to expect that reality may be more complicated than the simple scenario above. Models based on the described paradigm usually require that the energy produced by SN explosions (as well as by the stellar winds of young stars) must be transferred to the surrounding medium with an efficiency close to unity. Solutions must then be envisaged to stem the radiative losses that are expected on theoretical grounds (e.g.\ \citealt{Weaver1977}; \citealt{Gatto+17}; \citealt{NaabOstriker2017} and references therein). Even at maximum efficiency, it is widely believed that SN feedback is insufficient to explain the high-mass end of the galaxy stellar mass function, so that additional mechanisms are often advocated to regulate the evolution of galaxies in that regime, the most common being feedback from active galactic nuclei (AGN, e.g.\ \citealt{CiottiOstriker2001}; \citealt{BrighentiMathews2006}; \citealt{Beckmann+17}). AGN feedback has been modelled in a variety of fashions: thermal, kinetic, radiative, continuous or intermittent, isotropic or anisotropic, with a variety of assumed efficiencies and a different impact on the baryon content of the CGM (e.g. \citealt{Sijacki+07}; \citealt{Steinborn+15}; \citealt{Choi+15}; \citealt{Weinberger+17}). In addition to the above, other mechanisms are being investigated which can influence accretion on to galaxies and star formation, while having a variety of impacts on the baryon content of haloes. These include, among others, various kinds of radiation pressure (e.g.\ \citealt{Roskar+14}; \citealt{Kimm+18}), cosmic rays (e.g.\ \citealt{Samui+18}; \citealt{Jacob+18}), thermal conduction (\citealt{NipotiBinney2007}) and photo-ionization by local sources (\citealt{Cantalupo2010}; \citealt{Kannan+16}).

Observational determinations of the baryonic mass in the CGM around galaxies, as a function of halo mass and redshift, are of great help to confirm galaxy formation models or direct theoretical investigations further. At low redshift, a wealth of information comes from the X-ray bremsstrahlung of hot (close to virial temperature) gas in groups and clusters, as well as around galaxies (e.g.\ \citealt{Lagana+13}; \citealt{Bogdan+17}; \citealt{Li+18}). Some of these observations (e.g.\ \citealt{Lovisari+15}; \citealt{Eckert+16}) suggest that at $z = 0$ the fraction of baryons (at least of those in the hot phase) found within haloes increases with halo mass, an input that has already proven useful to calibrate feedback models in some cosmological simulations (e.g.\ \citealt{LeBrun+14}; \citealt{Pillepich+18}). In addition to observations of the hot gas, an increasing amount of data is becoming available for the colder phase ($T \sim 10^4 \; \textrm{K}$), entrained within hot haloes and mostly observable in absorption along the sightlines of background QSOs. Estimating the mass of this component is more difficult, because of the intrinsically serendipitous nature of absorption studies, as well as the need to resort to photo-ionization modelling to convert the observed column densities of the absorbing species into a total gas mass. It has however been suggested that the mass in the cold phase can in some cases also be significant (e.g.\ \citealt{Werk+14}; \citealt{Stern+16}; \citealt{Zahedy+19}).

The baryon content of the CGM of high $z$ galaxies is of special value to galaxy formation theory, as its determination would provide the most direct information on the nature of feedback mechanisms -- and in general on the processes determining galaxy formation -- in the earliest stages of their development. In particular, at early epochs, the amount of retained baryons is expected to change, with halo mass and redshift, in a way that is mostly determined by the efficiency of SN feedback (e.g.\ \citealt{Henriques+19}). Therefore, measurements of the CGM baryon fraction at high $z$ can, at least in principle, add useful constraints to the choice of the parameters, such as the fraction of SN energy that is converted into gas kinetic energy, that eventually modulate the ejection of baryons from galaxy haloes. Unfortunately, secure measurements of bremsstrahlung emission from hot haloes at high $z$ require very long integration times with the current instrumentation. Progress is expected from complementary probes, such as the Sunyaev-Zeldovich effect. This is unaffected by cosmological dimming and has already provided interesting indications at intermediate redshifts at the mass scale of galaxy clusters (e.g.\ \citealt{Chiu+18} and references therein). The mass of the hot gas can also be estimated indirectly from that of the cold gas, when observations of the latter are available, based on the assumption of pressure balance between the two phases. The \emph{ansatz} of pressure equilibrium was initially proposed by \cite{Spitzer1956}, who successfully applied it to the halo of the Milky Way, and was proven applicable up to the scale of galaxy clusters (e.g.\ \citealt{Churazov+01}; \citealt{Zhuravleva+16}). Absorption data provide some constraints on the cold gas around star-forming galaxies (\citealt{Rudie+12}) and QSOs (\citealt{Prochaska+13}) out to $z \sim 2$ and suggest the presence of significant amounts of cold gas at projected distances comparable to virial radius, though a quantitative determination of the total gas mass is not straightforward for the reasons mentioned above. On the other hand, the cold phase of the CGM at $z \gtrsim 2$ is sufficiently bright in fluorescent Ly$\alpha$ emission to be detected with current facilities, as long as it is illuminated by an intense source of ionizing photons, such as a bright QSO. Indeed, numerous extended Ly$\alpha$ nebulae associated with AGN have been discovered around radio galaxies (e.g.\ \citealt{VillarMartin+07} and references therein) and more recently around radio-quiet QSOs (e.g.\ \citealt{Cantalupo+14}; \citealt{Hennawi+15}).

Thanks to the Multi Unit Spectroscopic Explorer (MUSE; \citealt{Bacon+10}), \cite{Borisova+16} found 100\% detection rate of giant Ly$\alpha$ nebulae around a sample of 17 radio-quiet and 2 radio-loud very luminous ($L \sim 10^{44} \; \textrm{erg} \; \textrm{s}^{-1}$) QSOs at $z \sim 3$. A similar result was also found by the MUSEUM survey (\citealt{MUSEUM}). Giant Ly$\alpha$ nebulae around luminous radio-quiet QSOs are the ideal laboratory where to investigate the CGM baryon fraction, for at least two reasons. First, bright QSOs are able to ionize the illuminated part of their own CGM almost completely, so that in this region contributions to Ly$\alpha$ emission from collisional excitation and scattering either are drastically reduced or their treatment is much simpler than in more ordinary environments. Second, both the typical QSO life time and the CGM light crossing time are much shorter than the CGM sound crossing time. This gives us some confidence that fluorescent nebulae around radio-quiet QSOs provide a genuine and unbiased snapshot of the state of CGM, \emph{before} it might be significantly displaced by possible mechanical feedback from the current QSO activity event. \footnote{The same does not necessarily apply to radio-loud QSOs, as relativistic jets might displace some portions of the CGM as rapidly as the ionizing photons make them visible through fluorescence. Note, nonetheless, that the two nebulae discovered by \cite{Borisova+16} around radio-loud QSOs have a surface brightness profile very similar to that of nebulae around radio-quiet objects.}

The aim of this paper is to build upon the observations of \cite{Borisova+16} and provide an estimate of the CGM baryon fraction in relatively massive $(M_\textrm{h} \sim 10^{12} \; \textrm{M}_\odot)$ haloes at $z\sim3$. We find that a Ly$\alpha$ surface brightness as high as the observed one is difficult to explain with baryon fractions significantly smaller than the cosmological average. Lower values, typical of strong ejective feedback models, can be accounted for under specific conditions, which we discuss in detail and which can be tested by independent observations. Much of our analysis relies on the hypotheses that (i) Ly$\alpha$ emission is dominated by recombination radiation and (ii) that the cold phase and the hot phase are in pressure equilibrium with each other. We also discuss, however, the conditions and consequences of relaxing these assumptions.

In Section \ref{sec::coldCGM} we use the observations of \cite{Borisova+16} to infer the density profile and the total mass of the cold CGM gas in massive haloes at $z \sim 3$. In Section \ref{sec::Hot} we estimate the mass of the hot gas based on the assumption of pressure equilibrium. In Section \ref{sec::Tot} we discuss the total CGM mass, its dependence on model parameters and the conditions to reconcile it with the strong ejective feedback paradigm. In Section \ref{sec::ion} we check the self-consistency of our model in terms of the ionization structure of the CGM. Section \ref{sec::Scattering} addresses deviations from a pure-recombination scenario and in particular the impact of Ly$\alpha$ scattering. In Section \ref{sec::Discussion} we discuss the possible origin and physical properties of the cold gas, including possible deviations from pressure equilibrium. A summary is given in Section \ref{sec::Summary}.

In this paper we consistently adopt a flat cosmology with parameters $H_0 = 67.74 \; \textrm{km} \; \textrm{s}^{-1} \; \textrm{Mpc}^{-1}$, $\Omega_m = 0.3089$ and a cosmic baryon fraction $\Omega_b/\Omega_m = 0.1573$ (\citealt{Planck16}). Using cosmological parameters as in \cite{Borisova+16} has negligible impact on our results.

\section{The cold CGM in MUSE QSO Ly$\alpha$ nebulae}\label{sec::coldCGM}

\subsection{Recombination radiation from the CGM}

\subsubsection{Ly$\alpha$ recombination surface brightness}
If a galaxy hosts a sufficiently bright QSO, large portions of its CGM can be almost completely photo-ionized (see Section \ref{sec::ion} below for more details). Recombinations in the medium would produce Ly$\alpha$ radiation, with a surface brightness given by the integral along the line of sight of the recombination Ly$\alpha$ emissivity
\begin{equation}\label{SBintegral}
\Sigma = \int f_V B(T) n^2 \; ,
\end{equation}
where $f_V$ and $n$ are, respectively, the \emph{volume filling factor} and the hydrogen number density of the Ly$\alpha$ emitting gas, while the coefficient $B$ is given by
\begin{equation}\label{Bdef}
B(T) \equiv \frac{E_{\textrm{Ly}\alpha}}{4 \pi} \left( 1 + 2 \frac{n_\textrm{He}}{n_\textrm{H}} \right) \alpha^\textrm{eff}_{\textrm{Ly}\alpha}(T) \; ,
\end{equation}
where $E_{\textrm{Ly}\alpha}$ is the energy of the Ly$\alpha$ line, $n_\textrm{He}/n_\textrm{H} = 0.25 Y/(1-Y) = 1/12$ is the relative proportion of helium-to-hydrogen nuclei for a helium abundance by mass $Y = 0.25$ and finally (posing $T_4 \equiv T/10^4 \; \textrm{K}$)
\begin{equation}\label{alphaeff}
\alpha^\textrm{eff}_{\textrm{Ly}\alpha} = 1.6 \times 10^{-13} \; T_4^{-0.96} \; \textrm{cm}^3 \; \textrm{s}^{-1}
\end{equation}
is the effective Ly$\alpha$ recombination coefficient, as derived by fitting the calculations by \cite{Pengelly64} in the temperature range $5 \times 10^3 \; \textrm{K} < T < 8 \times 10^4 \; \textrm{K}$. We adopt here case A coefficients. Note however that case A and B \emph{effective} Ly$\alpha$ recombination coefficients differ by only $5 \%$ (\citealt{Pengelly64}). They must not be confused with the \emph{total} recombination coefficients, which differ in the two cases by a factor of $\sim 1.6$ (e.g.\ \citealt{OsterbrockFerland}).

\subsubsection{Filling factor}\label{subsubsec::fV}
In a multiphase CGM, hot and diffuse gas at $T \sim 10^6 - 10^7 \; \textrm{K}$ occupies the majority of the volume and compresses the relatively cold phase ($T \sim 1 - 5 \times 10^4 \; \textrm{K}$) to large densities and a small fraction of the volume. Given its dependence on density and temperature (equations \ref{SBintegral} and \ref{alphaeff}), recombination radiation will be dominated by the cold and dense gas for any filling factor larger than $\sim 10^{-6}$. This justifies the approximation that Ly$\alpha$ emission is produced by the cold phase only and explains the importance of including the filling factor $f_V$ in equation \eqref{SBintegral}.

We do not make assumptions here about the geometry or topology of the cold Ly$\alpha$ emitting regions (for instance, whether they are disconnected clouds or clumps, or a web of sheets and filaments), nor on their typical size, and we will loosely refer to them as \emph{cold gas structures}. The possible nature and origin of the cold gas structure, as well as their typical sizes, are discussed further in Section \ref{sec::Discussion}.

If not stated otherwise, we interchangeably indicate with $n_\textrm{cold}$ or simply with $n$ the number density of hydrogen nuclei \emph{within} the cold gas structures. This should not be confused with the \emph{volume-averaged} density of cold gas, which is given by
\begin{equation}\label{aven}
\langle n \rangle = f_V n \; ,
\end{equation}
where the average is performed on volumes small with respect to the nebula but large with respect to the size of the emitting structures. Note that the term $f_V n^2$ in equation \eqref{SBintegral} could equivalently be replaced by $C\langle n \rangle^2$, where
\begin{equation}\label{C}
C = f_V^{-1}
\end{equation}
is the \emph{clumping factor} of the emitting gas, as due to the fact that the latter only occupies a fraction of the total volume.

In this study, we neglect \emph{small-scale} variations of the density \emph{within} the structures themselves, i.e.\ their \emph{internal clumpiness}, which would appear as an additional multiplicative factor $C_\textrm{int}$ on the right-hand side of equation \eqref{C} (though see a discussion in Section \ref{subsec::overP} below and see also \citealt{Cantalupo+19}). We do, however, allow for \emph{large-scale} variations of the density $n$ with the physical distance $r$ from the centre of the nebula. This is both necessary -- to explain the observed radially declining surface brightness profile (Section \ref{subsec::ncoldprofile}) -- and expected, as the pressure of the confining hot gas is likely to decline with radius due to the gravitational pull of the halo (Section \ref{sec::Hot}). The filling factor $f_V$ and the temperature of the cold gas are also in principle a function of radius, though in the following we assume them to be constant, for simplicity.

\subsubsection{Main caveats}\label{subsec::caveats}
In the rest of this section, we apply equation \eqref{SBintegral} to the observed Ly$\alpha$ surface brightness profile of the MUSE QSO Giant Ly$\alpha$ nebulae, to infer the density and mass of the cold phase of the CGM around bright QSOs at $z\sim 3$. When doing so, one must be aware of a number of effects that could bias the result in one direction or the other. Dust absorption can attenuate the Ly$\alpha$ flux. We neglect dust, as we do not expect it to be present in the CGM or IGM in large amounts. Obviously, however, if the intrinsic Ly$\alpha$ surface brightness is larger than what we measure, our derived densities and masses should be interpreted as lower limits, strengthening our main finding of a large CGM baryon fraction. Similarly, if only a small fraction of the CGM is highly ionized (which may happen, for instance, because of a small aperture of the QSO ionization cone), then a significant portion of the existing cold CGM would not be visible in recombination radiation. This would also imply that our mass estimates are lower limits. On the other hand, mechanisms other than recombination, such as collisional excitation and scattering (or `photon pumping') can contribute to the observed surface brightness. Scattering within the CGM can also mix photons belonging to different lines of sight, complicating the analysis. The most important of these points are discussed in greater detail in Sections \ref{sec::ion} and \ref{sec::Scattering}. We argue there that they either do not affect our results, or can be taken into account by relatively simple modelling.

\subsubsection{A note on ionization by stars}\label{subsubsec::ionstars}
For some aspects of our problem -- in particular, to estimate the temperature of the emitting gas, which determines the value of the recombination coefficient (see equation \ref{alphaeff} and Section \ref{subsec::Tcold} below) -- it is important to determine what the main source of ionizing photons is.  In the close vicinity of a QSO, the QSO itself is obviously the most likely dominant source of ionizing photons. However, other sources such as young stars could in principle provide ionizing photons as well. For young stars to dominate the photoionization budget of the MUSE Giant Ly$\alpha$ nebulae, \emph{both} the following two conditions must be met: (i) that the ionization cone of the QSO is sufficiently narrow that most of the CGM is not illuminated by the QSO \emph{and} (ii) that there is sufficient star formation (either in the central galaxy or in the CGM itself) to keep the CGM completely ionized and power the observed Ly$\alpha$ luminosity. The first condition cannot be discarded, considering the large observational uncertainties (see e.g.\ \citealt{Ichikawa+19} and references therein), although recent work suggests that the most luminous QSOs may have rather wide ionization cones (\citealt{Ricci+17}). The second necessary condition, on the other hand, is disfavoured based on energetic grounds and observed scaling relations. To power the observed median luminosity $L_{\textrm{Ly}\alpha} = 10^{44} \; \textrm{erg} \; \textrm{s}^{-1}$, one would need a star formation rate of $100 \; \textrm{M}_\odot \; \textrm{yr}^{-1}$ (e.g.\ \citealt{DijkstraWestra2010}), assuming an implausibly large escape fraction of ionizing photons $f_\textrm{esc} = 1$, or $4 \times 10^3 \; \textrm{M}_\odot \; \textrm{yr}^{-1}$ assuming a more realistic value $f_\textrm{esc} \simeq 2.5 \%$ (e.g.\ \citealt{HM12} and references therein). For comparison, the observed stellar-to-halo mass relation (e.g.\ \citealt{Behroozi+13}; \citealt{Moster+13}) and main sequence of star-forming galaxies (e.g.\ \citealt{Speagle+14} and references therein) imply that a halo with mass $M_\textrm{halo} \simeq 10^{12} \; \textrm{M}_\odot$ (close to the peak of star formation efficiency) at $z\sim3$ would typically host a central galaxy with stellar mass $M_\star = 10^{10.4} \; \textrm{M}_\odot$ and a star formation rate of $80 \; \textrm{M}_\odot \; \textrm{yr}^{-1}$, i.e.\ 50 times smaller than required. Similarly, diffuse star formation within the CGM itself is an unlikely explanation for the ionization state of the MUSE nebulae, as it would require that the CGM forms stars at a rate equal to, or larger than, that in the central galaxy.

\subsection{De-projected cold gas density profile}\label{subsec::ncoldprofile}

The MUSE QSO Giant Ly$\alpha$ nebulae (\citealt{Borisova+16}) have a median intrinsic (redshift-dimming corrected) surface brightness profile that is well described by a power-law function of the projected radius $R$:
\begin{equation}\label{SBprofile}
\Sigma (R) = 3.9 \times 10^{-4} \left( \frac{R}{10 \; \textrm{kpc}} \right)^{-1.5} \; \textrm{erg} \; \textrm{s}^{-1} \; \textrm{cm}^{-2} \; \textrm{srd}^{-1} \; .
\end{equation}
Note that the sample of \cite{Borisova+16} contains 17 radio-quiet and 2 radio-loud QSOs. We used here the entire sample; excluding the two radio-loud QSOs has negligible impact on our results. Note that the slope in equation \eqref{SBprofile} is different from the slope reported in \cite{Borisova+16}. Our slope was obtained by fitting to the median of the individually dimming-corrected surface brightness profiles, improving on the rough first estimate in \cite{Borisova+16}. Our results are not sensitive to small changes in the slope of the surface brightness profile (see also Section \ref{subsec::InternalScattering} below). We did not attempt to model individual nebulae, mainly because a measurement of the baryon fraction requires an estimate of the halo mass and this is known only in a statistical sense (see Section \ref{subsec::HaloMass}).

Assuming spherical symmetry (or as long as we focus on spherically averaged properties), we can deproject equation \eqref{SBprofile} by means of equation \eqref{SBintegral}, to obtain the intrinsic radial profile of the density of the cold (Ly$\alpha$-emitting) gas:
\begin{equation}\label{densprofile}
n (r) = 0.15 \; f_V^{-1/2} T_4^{t/2} \left( \frac{r}{10 \; \textrm{kpc}} \right)^{-1.25} \; \textrm{cm}^{-3} \; ,
\end{equation}
where $r$ is the spherical radius and $t = 0.96$ is the exponent in relation \eqref{alphaeff}. More in detail, the normalization and slope $(n_0, \gamma)$ of the de-projected density profile (equation \ref{densprofile}) are obtained from the normalization and slope $(\Sigma_0, \beta)$ of the projected surface brightness profile (equation \ref{SBprofile}) by means of the relations
\begin{equation}\label{n0}
\left.\begin{array}{c}
n_0 = f_V^{-1/2} \left( \dfrac{\Sigma_0}{2B (T)\chi(1+\beta)R_0} \right)^{1/2} \; , \\
\\
\gamma = \dfrac{1+\beta}{2} \;,
\end{array}\right.
\end{equation}
where $R_0 = 10 \; \textrm{kpc}$ and the dimensionless projection factor $\chi$ is defined (see also \citealt{Pezzulli+17}) as
\begin{equation}\label{chidef}
\chi(a) \equiv \int_0^{+\infty} \left( 1 + x^2 \right)^{-a/2} d x \; .
\end{equation}
\begin{figure}
\centering
\includegraphics[width=9cm]{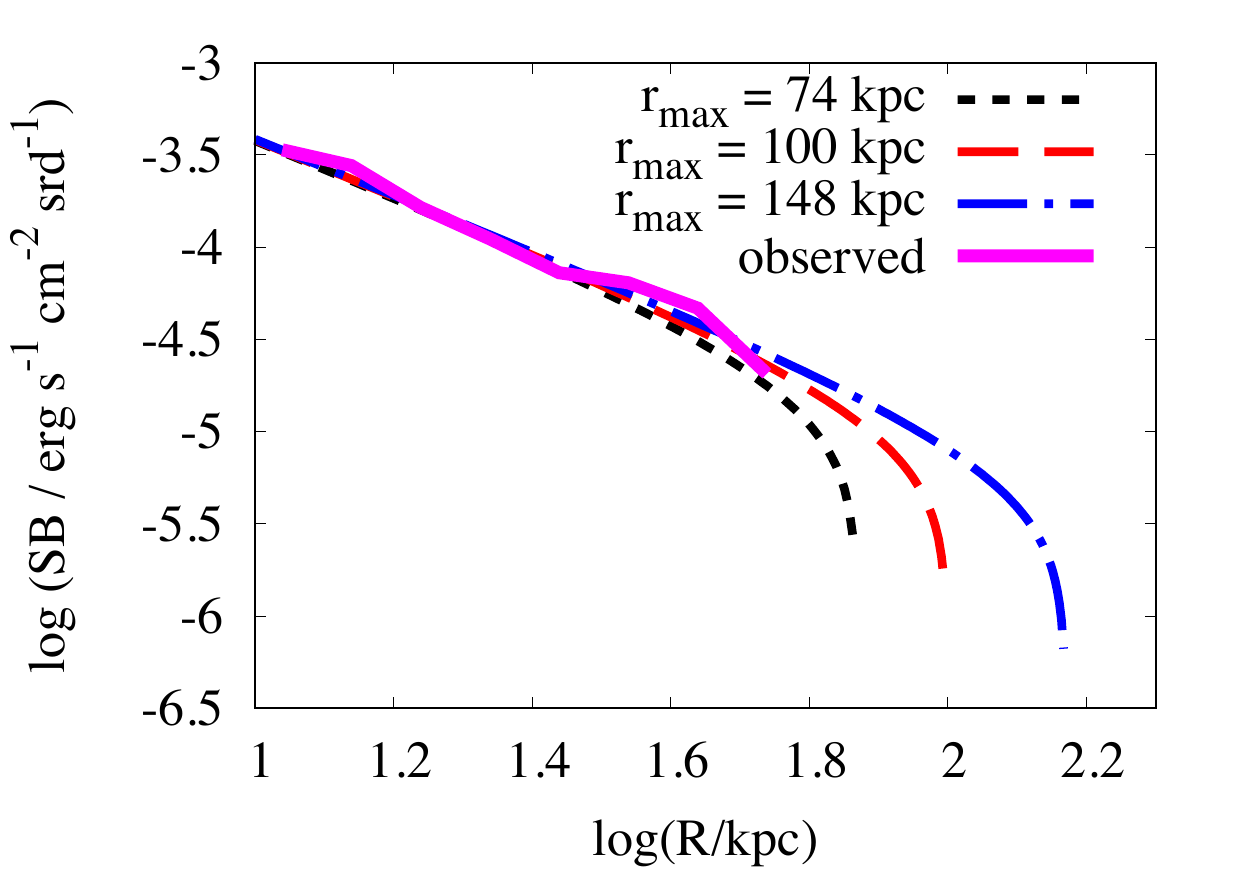}
\caption{Model Ly$\alpha$ surface brightness profile, for different choices of the truncation radius $r_\textrm{max}$, compared to the observed (redshift-dimming corrected) median profile of the MUSE Giant Ly$\alpha$ nebulae. The data do not constrain the exact value of $r_\textrm{max}$, but they support the assumption that it is equal to or larger than the virial radius (cf.\ equation \ref{Rvirdef}). \label{fig::SBtrunc}}
\end{figure}
The derivation above, which is equivalent to a classic Abel de-projection, formally relies on the assumption that the surface brightness profile (equation \ref{SBprofile}) and the density profile (equation \ref{densprofile}) extend to infinity. As this is not realistic, we show in Figure \ref{fig::SBtrunc} the effect, on the projected Ly$\alpha$ surface brightness profile, of introducing a truncation radius $r_\textrm{max}$ in the density profile \eqref{densprofile}. Predictions are shown for some values of $r_\textrm{max}$, equal to or larger than the virial radius of a $10^{12} \; \textrm{M}_\odot$ halo at $z = 3$ ($r_\textrm{vir} \simeq 74 \; \textrm{kpc}$, see equation \ref{Rvirdef} below). Truncation radii equal or larger than $r_\textrm{vir}$ are all consistent with the current data. Recent results from the MUSEUM survey (\citealt{MUSEUM}) possibly support a truncation radius not much larger than the virial radius. On the other hand (as already noted by \citealt{Borisova+16}), there is at least one example of a Giant Ly$\alpha$ nebula whose surface brightness profile is identical to those of the MUSE nebulae and extends to considerably larger physical radii (the Slug nebula; \citealt{Cantalupo+14}). We note, however, that azimuthal symmetry is a poor description of the Slug nebula, as well as of the outer regions of some of those MUSE nebulae that can be traced at radii larger than $r_\textrm{vir}$, so that the interpretation of profiles extending beyond the virial radius is not obvious. In the following, we will conservatively assume that the profile \eqref{densprofile} holds within the virial radius, which is what we need to estimate the CGM baryon fraction there. In any case, we emphasize that our analysis is not meant to be applied far out in the IGM, where both strong asymmetries and lack of pressure confinement from a hot atmosphere are expected.

The dependencies in equation \eqref{densprofile} on $f_V$ and $T_4$ are readily understood: if a smaller fraction of the volume is occupied by Ly$\alpha$ emitting gas, individual structures must be denser in order to explain the observed emission; at the same time, a larger temperature implies a smaller effective recombination coefficient (equation \ref{alphaeff}) and therefore again a larger density to give account of a fixed Ly$\alpha$ surface brightness. Note that the relation between inferred density and filling factor in equation \eqref{densprofile} is equivalent to the familiar relation between inferred \emph{average} density and clumping factor $\langle n \rangle \; \propto \; C^{-1/2}$, as it follows from equations \eqref{aven} and \eqref{C}.

\subsection{The baryon fraction of the cold CGM}\label{subsec::fbcold}

The density profile \eqref{densprofile} can be integrated over volume to obtain the total mass of cold gas within the virial radius:
\begin{equation}\label{Mcold}
M_\textrm{cold} = \int_0^{r_\textrm{vir}} 4 \pi f_V \frac{m_p}{X} r^2 n(r) dr \; ,
\end{equation}
where $m_p$ is the mass of the proton and $X = 0.75$ is the cosmic hydrogen abundance by mass, while, as customary, we approximate the virial radius with the radius $r_{200}$ enclosing a 200-fold matter overdensity:
\begin{equation}\label{Rvirdef}
r_\textrm{vir} = \dfrac{1}{1+z} \left( \dfrac{G M_\textrm{halo}}{100 \; \Omega_\textrm{m} H_0^2} \right)^{1/3} = 74 \; M_{12}^{1/3} \; \textrm{kpc} \; ,
\end{equation}
where $M_{12} \equiv M_\textrm{halo}/10^{12} \; \textrm{M}_\odot$ is the halo mass in units of $10^{12} \; \textrm{M}_\odot$ and $z = 3.2$ is the median redshift of the MUSE QSO Ly$\alpha$ nebulae. Finally, dividing $M_\textrm{cold}$ by the cosmological baryon fraction gives the fraction, among all the baryons nominally associated with the  halo, that are stored in the cold phase of the CGM:
\begin{equation}\label{mucold}
\left. \begin{array}{ll}
f_\textrm{CGM,cold} & \equiv \dfrac{M_{\textrm{cold}}(r_\textrm{vir})}{(\Omega_b/\Omega_m) M_\textrm{halo}} \\
&\\
&= f_1 \; f_V^{1/2} T_4^{t/2} M_{12}^{-\gamma/3} \; ,
\end{array}\right.
\end{equation}
where $f_1 = 7.3$, $t = 0.96$ and $\gamma = 1.25$.  

As $f_\textrm{CGM,cold}$ cannot exceed unity, equation \eqref{mucold} implies that the filling factor of the cold gas $f_V$ must be smaller than 1 (or, equivalently, that the clumping factor must be larger than 1, cfr.\ equation \ref{C}), in agreement with previous findings (e.g.\ \citealt{Cantalupo+14}; \citealt{FAB15}). We now consider the option that the physical reason why the cold gas has $f_V < 1$ is pressure confinement by a hot corona, which is likely present in the halo, although invisible in Ly$\alpha$ emission.

\section{The mass of the hot CGM}\label{sec::Hot} 

Here we calculate the mass of the hot phase of the CGM that is needed to confine the cold phase to the densities derived in Section \ref{sec::coldCGM}. Our treatment is based on the assumption of pressure equilibrium between the two phases and is not dissimilar, in its principles, to other works on the subject (e.g.\ \citealt{HaimanRees2001}; \citealt{DijkstraLoeb2009}; \citealt{VedanthamPhinney2019}). The possible presence and impact of deviations from pressure equilibrium are briefly discussed at the end of this section and in more detail in Section \ref{sec::Discussion}.

Pressure equilibrium dictates that the ratio $\varepsilon$ of the densities of the two phases (i.e.\ the \emph{density contrast}) equals the inverse ratio of their temperatures:
\begin{equation}\label{Peq}
\varepsilon = \frac{n_\textrm{hot}}{n_\textrm{cold}} = \frac{T_\textrm{cold}}{T_\textrm{hot}} \; ,
\end{equation}
where $n_\textrm{hot}$ and $T_\textrm{hot}$ are the hydrogen number density and the temperature of the hot gas, respectively. Note that, in equation \eqref{Peq}, we do not need to account for different ionized fractions (and therefore a different mean molecular weight) for the cold gas and the hot gas, as both phases are assumed here to be virtually completely ionized (although by entirely different processes: collisional ionization and photo-ionization for the hot and cold phases, respectively). 

If one assumes that both $T_\textrm{cold}$ and $T_\textrm{hot}$ are independent on radius, then the density contrast $\varepsilon$ is also a constant, with the consequence that the inferred density profile of the hot gas follows a similar power law as the cold gas (equation \ref{densprofile}), with a downward normalization shift, due to the fact that the hot gas is more diffuse:
\begin{equation}\label{powerlawhot}
n_\textrm{hot}(r) = \varepsilon n_0 \left( \frac{r}{R_0} \right)^{-\gamma} \;.
\end{equation}
Interestingly, a power-law profile like that in equation \eqref{powerlawhot} is indeed consistent with \emph{isothermal} hydrostatic equilibrium in a dark matter halo gravitational potential, for a gas with close-to-virial temperature
\begin{equation}\label{Thot}
T_\textrm{hot} = \frac{1}{\gamma} \frac{\mu m_p}{k_B} \frac{GM_\textrm{halo}}{r_\textrm{vir}} \; ,
\end{equation}
where $m_p$ is the mass of the proton, $k_B$ the Boltzmann constant and $\mu = 0.593$ the mean molecular weight. This suggests that temperature gradients, if present, are not dramatic.

By combining equations \eqref{Rvirdef}, \eqref{Peq} and \eqref{Thot}, the density contrast can be written
\begin{equation}\label{epsilon}
\begin{split}
\varepsilon & = \frac{\gamma k_B T_\textrm{cold}}{\mu m_p} \left( 1 + z \right)^{-1} \left( 100 \; \Omega_\textrm{m} H_0^2 \right)^{-1/3} \left( G M_\textrm{halo} \right)^{-2/3} \\
& = \varepsilon_1 T_4 M_{12}^{-2/3} \; ,
\end{split}
\end{equation}
with
\begin{equation}\label{epsilon1}
\varepsilon_1 = 1.0 \times 10^{-2} \frac{\gamma}{1 + z} = 3.0 \times 10^{-3}
\end{equation}
for $z = 3.2$ and $\gamma = 1.25$.

The total mass of the hot gas within the virial radius is readily obtained by integrating equation \eqref{powerlawhot}. Equivalently, the ratio of hot to cold gas mass is given by the ratio of the densities times the ratio of the volumes occupied by the two phases 
\begin{equation}\label{hotcoldratio}
\frac{M_\textrm{hot}}{M_\textrm{cold}} = \varepsilon \frac{1 - f_V}{f_V} = \varepsilon_1 f_V^{-1} T_4 M_{12}^{-2/3} \;,
\end{equation}
where we have used equation \eqref{epsilon} and taken into account that $f_V \ll 1$ for every viable model (cf.\ Section \ref{subsec::fbcold}). 

The quantitative implications of equations \eqref{mucold} and \eqref{hotcoldratio} are presented in Section \ref{sec::Tot}. Before doing so, we briefly discuss here whether any significantly large errors on the mass of the hot gas are expected to be introduced by our simplifying assumptions.

It is easy to verify that the simple inference above is not affected by the detailed shape of the gravitational potential. In a realistic potential (e.g.\ \citealt{NFW}), the hydrostatic equilibrium density profile of an isothermal hot gas would not be \emph{exactly} described by a power law, but rather require an additional radial-dependent factor equal to $(V_\textrm{circ}(r)/V_{200})^{-2}$, where $V_\textrm{circ} (r)$ is the circular speed at radius $r$ and $V_{200} = V_\textrm{circ} (r_\textrm{vir})$. In the situation of interest, this correction is remarkably small (less than 10\% for $0.2 < r/r_\textrm{vir} < 1$), mostly due to the relatively low concentration of dark matter haloes at this redshift ($c \sim 3.5$; e.g.\ \citealt{DuttonMaccio2014}). The impact on the total mass is even smaller (less than 1\%), mainly because most of the mass is at large radii, where the approximation $V_\textrm{circ} \simeq V_{200}$ is the most accurate. Coherent rotation, which must be present as the CGM necessarily has non-negligible angular momentum, also affects the density profile by providing centrifugal support against gravity, but also in this case the impact is mostly limited to the very innermost regions, which contain a very small fraction of the mass (\citealt{Pezzulli+17}; \citealt{Oppenheimer2018}).

More than the detailed shape of the potential or the presence of coherent rotation, deviations from the simple scenario above can arise due to large-scale turbulence, or other sources of non-thermal pressure. Crucially, however, such deviations would contribute \emph{both} to enhance the compression of the cold phase \emph{and} to provide additional support to the hot gas against gravity, with the result that the density contrast $\varepsilon$ (which is essentially estimated by taking the ratio of the two effects) would remain largely unaffected. A formal way to see this is to reformulate equation \eqref{Peq} in terms of a ratio of sound speeds squared $\varepsilon = c^2_{s, cold}/c^2_{s,hot}$. If large-scale turbulence is in place, the cold gas would be subject to (and compressed by) additional \emph{ram pressure}, which would be, at least to the first order, encapsulated by the replacement $c^2_\textrm{s,hot} \rightarrow c^2_\textrm{s,hot} + v^2_\textrm{turb, hot}$. An analogous reformulation and replacement would be needed in equation \eqref{Thot} as well, as the very same turbulent motions would also provide support to the hot gas against gravity.\footnote{viz., the equation $c_\textrm{s, hot}^2 = G M_\textrm{halo}/\gamma r_\textrm{vir}$ (equivalent to equation \ref{Thot}) should be replaced by $(c_\textrm{s, hot}^2 + v_\textrm{turb,hot}^2) = G M_\textrm{halo}/\gamma r_\textrm{vir}$. Note that only here we are using the `isothermal sound speed' $c_s \equiv \sqrt{P/\rho}$ to simplify the notation and without loss of generality.} It is then easily seen that the final estimated contrast $\varepsilon$ remains unaffected.\footnote{The same cannot be said for the temperature of the hot gas and therefore, for instance, for predictions of the X-ray luminosity of the CGM.} Although we cannot claim this result to be \emph{exact}, physical insight leads us to believe that additional corrections, if present, should be second-order and sub-dominant. Similar arguments should hold essentially for any \emph{large-scale} source of non-thermal pressure. Considerations about other kinds of pressure imbalance are discussed in Section \ref{sec::Discussion}.

\section{The total CGM baryon fraction}\label{sec::Tot}

By combining equations \eqref{mucold} and \eqref{hotcoldratio}, we finally obtain the \emph{total} (cold plus hot) CGM baryon fraction within the virial radius:
\begin{equation}\label{mutot}
\left.\begin{array}{ll}
f_\textrm{CGM,tot} & \equiv \dfrac{M_{\textrm{cold}+\textrm{hot}}(r_\textrm{vir})}{(\Omega_b/\Omega_m) M_\textrm{halo}} \\
&\\
&= f_1 ( 1 + \varepsilon_1 f_V^{-1} T_4 M_{12}^{-2/3} ) f_V^{1/2} T_4^{t/2} M_{12}^{-\gamma/3} \; ,
\end{array}\right.
\end{equation}
where we recall that $f_1 = 7.3$, $\varepsilon_1 = 3.0 \times 10^{-3}$, $t = 0.96$ and $\gamma = 1.25$ (equations \ref{alphaeff}, \ref{n0}, \ref{mucold} and \ref{epsilon1}).

The total CGM baryon fraction thus depends on three parameters: $f_V$, $M_\textrm{halo}$ and $T_\textrm{cold}$. We discuss these dependencies below, one at a time.

\subsection{Dependence on the filling factor}\label{subsec::fillingfactor}
\begin{figure} 
\centering
\includegraphics[width=9cm]{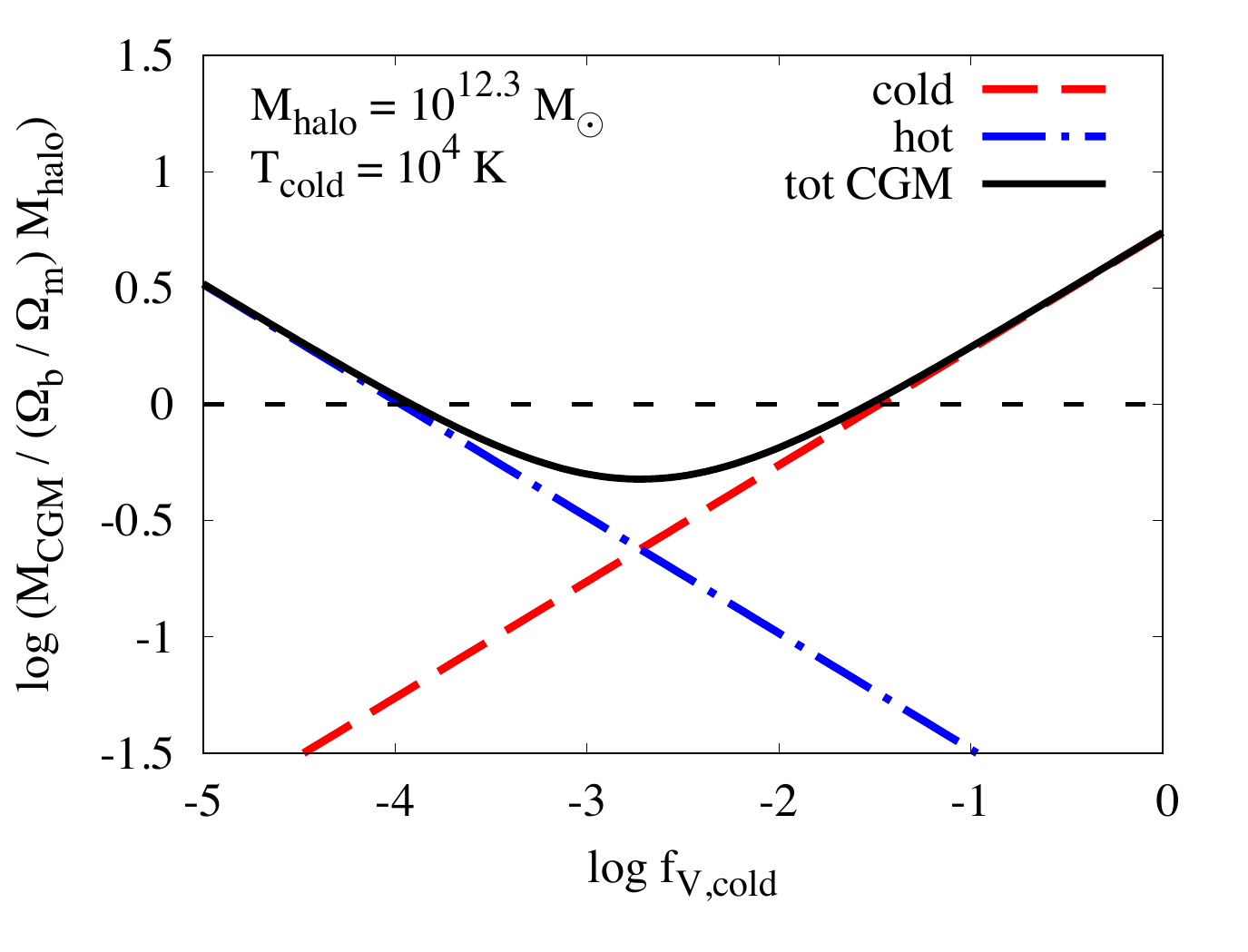}
\caption{Cold (red, long dashed), hot (blue, dot-dashed) and total (black, solid) mass of the CGM in our model of the MUSE Giant Ly$\alpha$ nebulae, in units of the maximum possible baryonic mass given by the cosmological baryon fraction (horizontal dashed line), as a function of the filling factor of the cold gas $f_V \equiv f_{V,cold}$. A halo mass $M_\textrm{halo} = 10^{12.3} \; \textrm{M}_\odot$ and a cold gas temperature $T = 10^4 \; \textrm{K}$ are fixed for illustrative purposes (see Figures \ref{fig::Mplot} and \ref{fig::Tplot} for other choices). The inferred masses of cold gas and hot gas have opposite dependencies on $f_V$, with the consequence that only a finite range of $f_V$ is allowed. Also, there is a \emph{minimum} CGM mass consistent with observations (47\% of the cosmological value for this choice of $M_\textrm{halo}$ and $T_\textrm{cold}$), which is achieved if the cold gas and the hot gas contribute equally to the total mass.
\label{fig::totplot}}
\end{figure}

In Figure \ref{fig::totplot} we show how the total CGM baryon fraction, inferred from MUSE observations and from our model, depends on the filling factor of the cold gas $f_V$. We adopt here a halo mass  $M_\textrm{halo} = 10^{12.3} \; \textrm{M}_\odot$ and a temperature of the cold gas $T_\textrm{cold} = 10^4 \; \textrm{K}$, but we stress that these values are fixed here for illustrative purposes only. The effects of varying $M_\textrm{halo}$ and $T_\textrm{cold}$ are discussed in detail in Sections \ref{subsec::HaloMass} and \ref{subsec::Tcold}, respectively. In Figure \ref{fig::totplot}, we also show the individual contributions of the cold gas and the hot gas to the total baryon fraction of the CGM.

As it is clear from the figure and from equation \eqref{mucold}, the inferred masses of the cold gas and the hot gas have opposite dependencies on $f_V$. The estimated mass of the cold CGM increases with increasing $f_V$. This is a consequence of the dependence of the recombination emissivity on the density squared (see Section \ref{sec::coldCGM}). On the other hand, the estimated mass of the hot gas increases with \emph{decreasing} $f_V$. This is mostly because a larger amount of hot gas is needed to compress the cold gas to the larger densities that are associated with smaller filling factors (Section \ref{sec::Hot}).

These opposite dependencies have two obvious consequences. First, there is at most a finite range of filling factors that are consistent with a physically meaningful baryon fraction (equal to or smaller than the cosmological value, shown as a horizontal black dashed line). Second, for any given choice of the other two parameters ($M_\textrm{halo}$ and $T_\textrm{cold}$), there is a minimum value of the total CGM baryon, below which no model based on recombination radiation and pressure balance would be able to explain the observed Ly$\alpha$ surface brightness of the MUSE nebulae. We will use this minimum in Section \ref{subsec::minfbtot} to put a \emph{lower limit} to the total CGM baryon fraction, as inferred from the MUSE observations within our assumptions. We emphasize that the minimum CGM baryon fraction is obtained when the two phases contribute equally to the total mass: if either of the two phases dominates in mass, the total baryon fraction will be larger than this lower limit.

\subsection{Dependence on halo mass}\label{subsec::HaloMass}
\begin{figure}
\centering
\includegraphics[width=9cm]{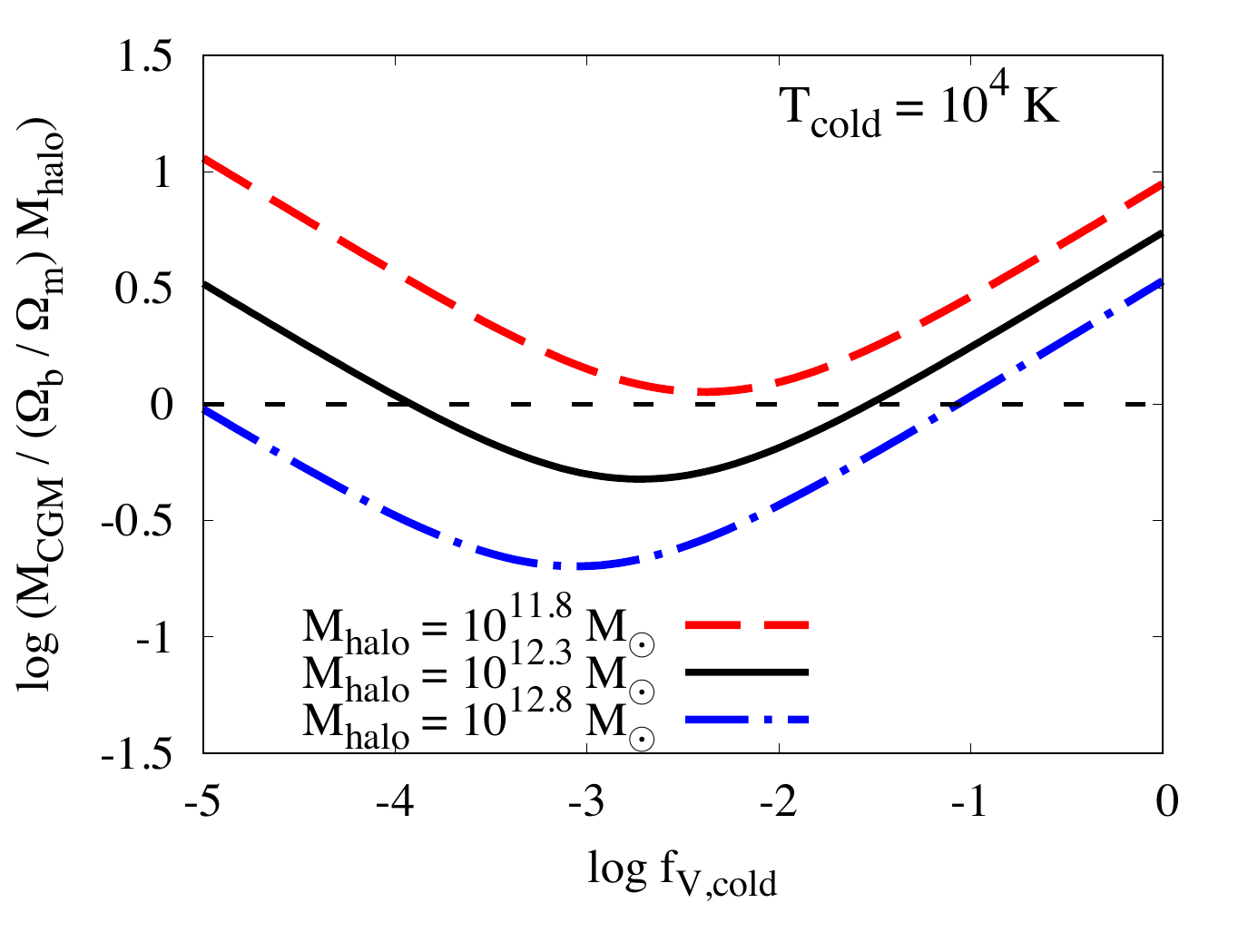}
\caption{Dependence of the total CGM baryon fraction on the filling factor of the cold gas, for different values of the assumed halo mass and a fixed temperature of the cold gas $T_\textrm{cold} = 10^4 \; \textrm{K}$. Results are shown for halo masses within the range inferred from galaxy-QSO cross-correlation (\citealt{TrainorSteidel2012}). Relatively low halo masses ($M_\textrm{halo} \sim 10^{12} \; \textrm{M}_\odot$), as recently suggested by BOSS QSO clustering (\citealt{Eftekharzadeh+15}), would imply a CGM baryon fraction close to the cosmological value. \label{fig::Mplot}}
\end{figure}

Figure \ref{fig::Mplot} shows the dependence of the inferred CGM baryon fraction on the assumed mass of the halo. Different lines correspond to the total (cold plus hot) CGM baryon fraction, for different values of the assumed halo mass and for one fixed value of the temperature of the cold gas $T_\textrm{cold} = 10^4 \; \textrm{K}$.

The inferred CGM baryon fraction decreases with increasing assumed halo mass. This is partly due to the obvious fact that increasing the halo mass also increases the amount of baryons theoretically associated with the halo, so that a lower fraction of those is needed to explain the observed Ly$\alpha$ surface brightness. Note however that the virial radius and the virial temperature are also changing accordingly, which explains why the dependence on halo mass is not trivially linear (equation \ref{mutot}).

With decreasing halo mass, it becomes increasingly difficult to accommodate a total CGM baryon fraction significantly smaller than the cosmological value, until eventually no physically acceptable solution is found. This makes the knowledge of the typical mass of QSO-hosting haloes at $z\sim3$ crucial to draw conclusions from our results.

Probably the most reliable estimates of QSO-hosting halo masses are those based on QSO clustering, or on the QSO-galaxy cross-correlation function. Several studies of this kind have been attempted at a redshift $z\sim3$ relevant to our application, finding a broad range of results, from $M_\textrm{halo} \sim 10^{12} \; \textrm{M}_\odot$ to $M_\textrm{halo} \sim 10^{13} \; \textrm{M}_\odot$ (see e.g.\ \citealt{Shen+07} and references therein). Big uncertainties are to a large extent due to low-number statistics, inherent to studies of rare objects such as QSOs, but possibly also to differences in selection techniques and especially in the methods used to determine the redshift of the sources, which is crucial for clustering measurements. Taking these aspects into account, we chose to use, as a reference, the study by \cite{Eftekharzadeh+15}. Their sample in fact is -- by far -- the one with the cleanest selection and the better statistics to our knowledge at the redshift of interest, with more than 20 000 QSOs with spectroscopic redshifts in the range $2.64 < z < 3.4$, as determined from the final sample of the Baryon Oscillation Spectroscopic Survey (BOSS; \citealt{Dawson+13_BOSS}). These authors find a typical halo mass $M_\textrm{halo} = 10^{12} \; \textrm{M}_\odot$, which is on the low side of the range of masses mentioned above. We refer to \cite{Eftekharzadeh+15} for a discussion of why larger values had been found by the previous work of \cite{Shen+07} on a smaller sample.

Despite the lack of new studies with sample sizes comparable to that of \cite{Eftekharzadeh+15}, the debate on the mass of QSO-hosting haloes at $z\sim3$ is still ongoing in the recent literature. For instance, \cite{Allevato+16} inferred a large mass $M_\textrm{halo} \sim 10^{12.9} \; \textrm{M}_\odot$ based on a sample of $\sim$ 400 X-ray selected AGN in a range $2.9 < z < 5.5$, $\sim 100$ of which with spectroscopic redshift. Also, \cite{Timlin+18} determined halo masses with a large uncertainty $M_\textrm{halo} = 1.70-9.83 \times 10^{12} \; \textrm{h}^{-1} \; \textrm{M}_\odot$ based on a sample of 1378 objects, identified as QSO candidates and assigned a photometric redshift in the range $2.9 < z < 5.1$ by a machine-learning algorithm. The results by \cite{Allevato+16} and \cite{Timlin+18} emphasize the importance of further investigations to attenuate uncertainties in this field. However, given the much smaller samples of these studies with respect to \cite{Eftekharzadeh+15}, as well as the less robust selection techniques (especially concerning the determination of redshift), we think that the BOSS result should, for the time being, be considered as the reference.

Even if the BOSS result is (at least formally) the best measurement available to date, it is still not obvious that it can be safely applied to our study. The sample of \cite{Eftekharzadeh+15} in the relevant redshift range contains QSOs as luminous as $\log(L_\textrm{bol} / \textrm{erg} \; \textrm{s}^{-1}) = 47.72$. This overlaps with the luminosity range of the QSOs in the sample of \cite{Borisova+16} ($47.29 < \log(L_\textrm{bol} / \textrm{erg} \; \textrm{s}^{-1}) < 48.14$, using equation 1 in \citealt{Shen+09}), but the latter also contains objects up to 2.6 times more luminous than the maximum luminosity probed by \cite{Eftekharzadeh+15}. If, as some theoretical models predict, the halo mass correlates strongly with QSO luminosities (see e.g.\ \citealt{Hopkins+07}; \citealt{ConroyMartin2013}), it is possible that the QSOs of \cite{Borisova+16} live in haloes with a larger mass than the QSOs in the BOSS sample. On the other hand, a number of studies have recently shown that statistical measurements of QSO host halo masses correlate surprisingly weakly with QSO luminosities (e.g.\ \citealt{Shen+09}; \citealt{Padmanabhan+09}; \citealt{Eftekharzadeh+15}; \citealt{Uchiyama+18}; \citealt{He+18}), so that this option is not very clearly supported by the current data.

Furthermore, we notice that the MUSEUM survey (\citealt{MUSEUM}) has investigated Giant Ly$\alpha$ nebulae around a sample of QSOs less luminous than those of \cite{Borisova+16}, reaching a maximum luminosity of $\log(L_\textrm{bol} / \textrm{erg} \; \textrm{s}^{-1}) = 47.87$. This is very similar to the maximum luminosity probed by \cite{Eftekharzadeh+15}, possibly making comparisons less prone to potential bias. The Giant Ly$\alpha$ nebulae in the MUSEUM sample have a surface brightness profile that is undistinguishable from those of \cite{Borisova+16} in the radial range of interest for this study, possibly suggesting that a bias in QSO luminosity is not the main explanation of the large surface brightnesses observed in all these objects, although studying Ly$\alpha$ nebulae around QSOs with luminosities even lower than those in the MUSEUM survey is needed to achieve a more complete picture on this matter.

Based on the discussion above, in the following, we will sometimes use as fiducial the halo mass $M_\textrm{halo} = 10^{12} \; \textrm{M}_\odot$ found by \cite{Eftekharzadeh+15}. However, we will also show and discuss results obtained in a broad range of masses $M_\textrm{halo} = 10^{12.3 \pm 0.5} \; \textrm{M}_\odot$ (\citealt{TrainorSteidel2012}), which we chose as representative of the range of uncertainty in the value of the halo mass.

If a low value $M_\textrm{halo} \sim 10^{12} \; \textrm{M}_\odot$ is confirmed by future studies, it would imply that it is not easy to explain the surface brightness of MUSE Giant Ly$\alpha$ nebulae with a baryon fraction significantly smaller than cosmological, at least within the assumptions discussed so far.

On the other hand, a significantly larger value for the halo mass would not \emph{necessarily} imply that the observed Ly$\alpha$ surface brightness is trivial to reproduce. Most models in fact agree that at sufficiently large halo masses most of the gas is expected to be in the hot phase. This would violate the first condition to maximize the predicted Ly$\alpha$ surface brightness and specifically that the two phases contribute equally to the total mass of the CGM (Section \ref{subsec::fillingfactor}). Equivalently, the relevant curve in Figure \ref{fig::Mplot} would need to be read at some (model-dependent) displacement from the minimum. The quantitative details of the transition can be rather different in different models, depending on both numerical aspects (e.g.\ \citealt{Nelson+13}) and on the adopted feedback recipe (e.g.\ \citealt{Correa+18}).

\subsection{Dependence on the temperature of the cold gas}\label{subsec::Tcold}

\begin{figure}
\centering
\includegraphics[width=9cm]{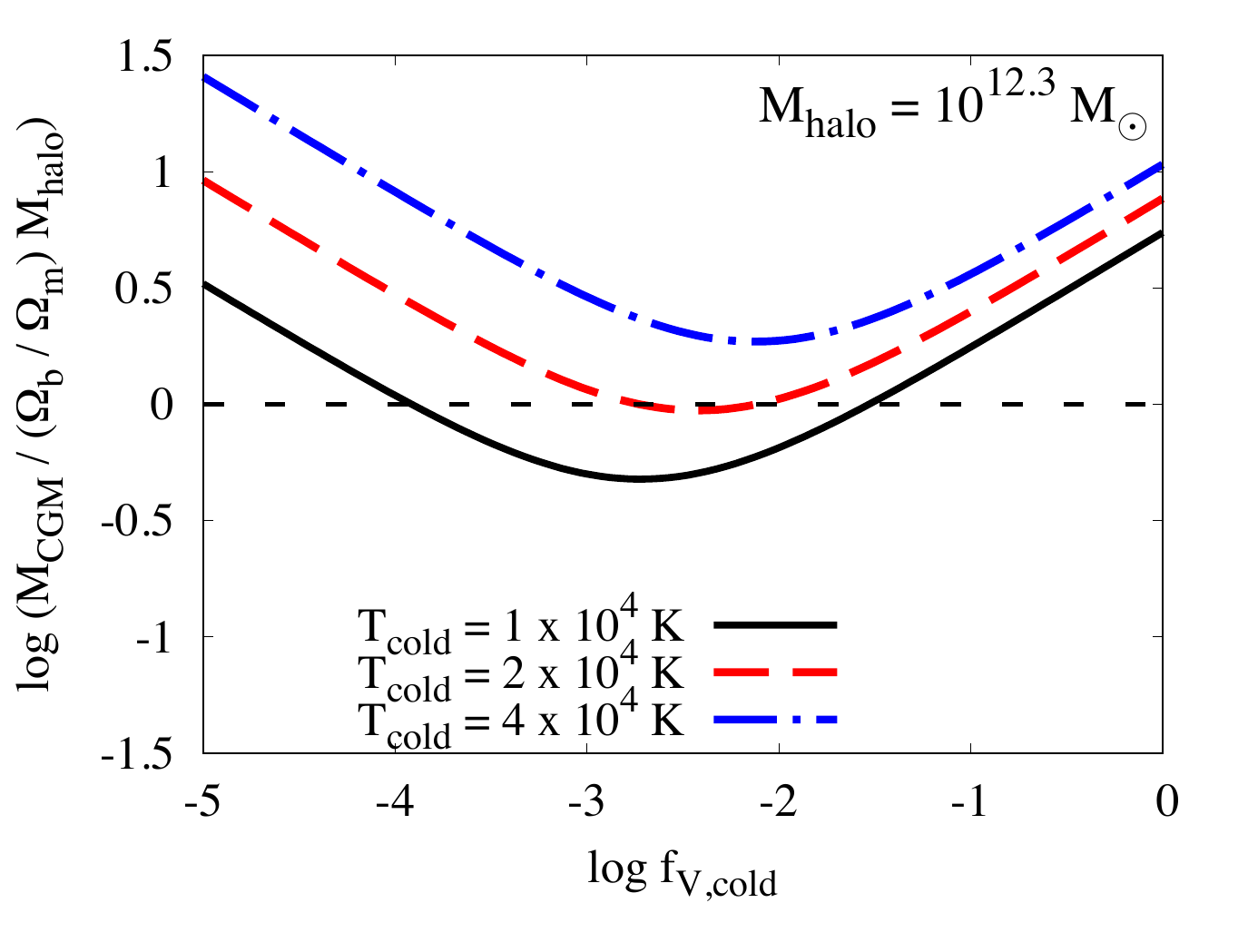}
\caption{Similar to Figure \ref{fig::Mplot}, but for a fixed halo mass $M_\textrm{halo} = 10^{12.3} \; \textrm{M}_\odot$ and different values of the temperature of the cold gas. A baryon fraction smaller than cosmological requires rather low temperatures, which are, however, difficult to achieve in the photo-ionization field of a QSO (see Figure \ref{fig::Teq} and the text for details). \label{fig::Tplot}}
\end{figure}

In Figure \ref{fig::Tplot} we show the dependence of the inferred total CGM baryon fraction on the assumed temperature of the cold photo-ionized gas $T_\textrm{cold}$, for a fixed halo mass $M_\textrm{halo} = 10^{12.3} \; \textrm{M}_\odot$. The inferred CGM baryon fraction increases with increasing the assumed $T_\textrm{cold}$. The reason is two-fold. First, as the recombination coefficient is a declining function of temperature (equation \ref{alphaeff}), a larger $T_\textrm{cold}$ requires a larger amount of cold gas to explain the same observed Ly$\alpha$ emission. Second, a larger $T_\textrm{cold}$ also implies a larger pressure of the cold gas and therefore a larger amount of hot gas needed to keep it confined. The magnitude of the effect is significant and it is therefore important to make physically motivated assumptions concerning the value of $T_\textrm{cold}$.

In an almost fully photo-ionized gas, the temperature is mainly determined by the equilibrium between photo-electric heating and recombination cooling and therefore, ultimately, by the shape of the ionizing spectrum (e.g.\ \citealt{OsterbrockFerland}). In our case, the ionizing spectrum is completely dominated by the QSO (we will see in Section \ref{sec::ion} that the UVB can be neglected by orders of magnitude). In a pure hydrogen nebula, the resultant equilibrium temperature would be
\begin{equation}
T_\textrm{eq} = \frac{2 h \nu_\textrm{LL}}{3 (p+2) k_B} = 2.8 \times 10^4 \left( \frac{p+2}{3.7} \right)^{-1} \; \textrm{K} \; ,
\end{equation}
where $h$ is the Planck's constant, $\nu_\textrm{LL}$ is the Lyman-limit frequency and $p$ is the power-law slope of the hydrogen ionizing continuum, for which we take $p = 1.7$ as a fiducial value (\citealt{Lusso+15}, see also Section \ref{sec::ion} below). 

In a QSO ionizing field, however, the photo-electric effect due to the second ionization of helium is also a very important heating channel. Taking this into account and assuming that a single power-law spectrum can be extrapolated to $E > 4 \; \textrm{Ry}$, the equilibrium temperature increases by a factor $(1 + 4 A(T))/(1 + A(T))$, where 
\begin{equation}
A(T) \equiv \frac{n_{He}}{n_H} \frac{\alpha^{He}_\textrm{rec}(T)}{\alpha^H_\textrm{rec}(T)} \; .
\end{equation}
By using the temperature-dependent recombination coefficients as in \cite{Meiksin2009}, we then find an equilibrium temperature
\begin{equation}\label{Teq}
T_\textrm{eq} = 5.6 \times 10^4 \; \textrm{K} \; .
\end{equation}
This is large enough that no physically acceptable solution can be found for a halo mass $M_\textrm{halo} = 10^{12.3} \; \textrm{M}_\odot$, as it is clear for instance from Figure \ref{fig::Tplot}. Note that the problem would only be exacerbated for $M_\textrm{halo} = 10^{12} \; \textrm{M}_\odot$, as suggested by clustering measurements (Section \ref{subsec::HaloMass}).

\begin{figure}
\centering
\includegraphics[width=9cm]{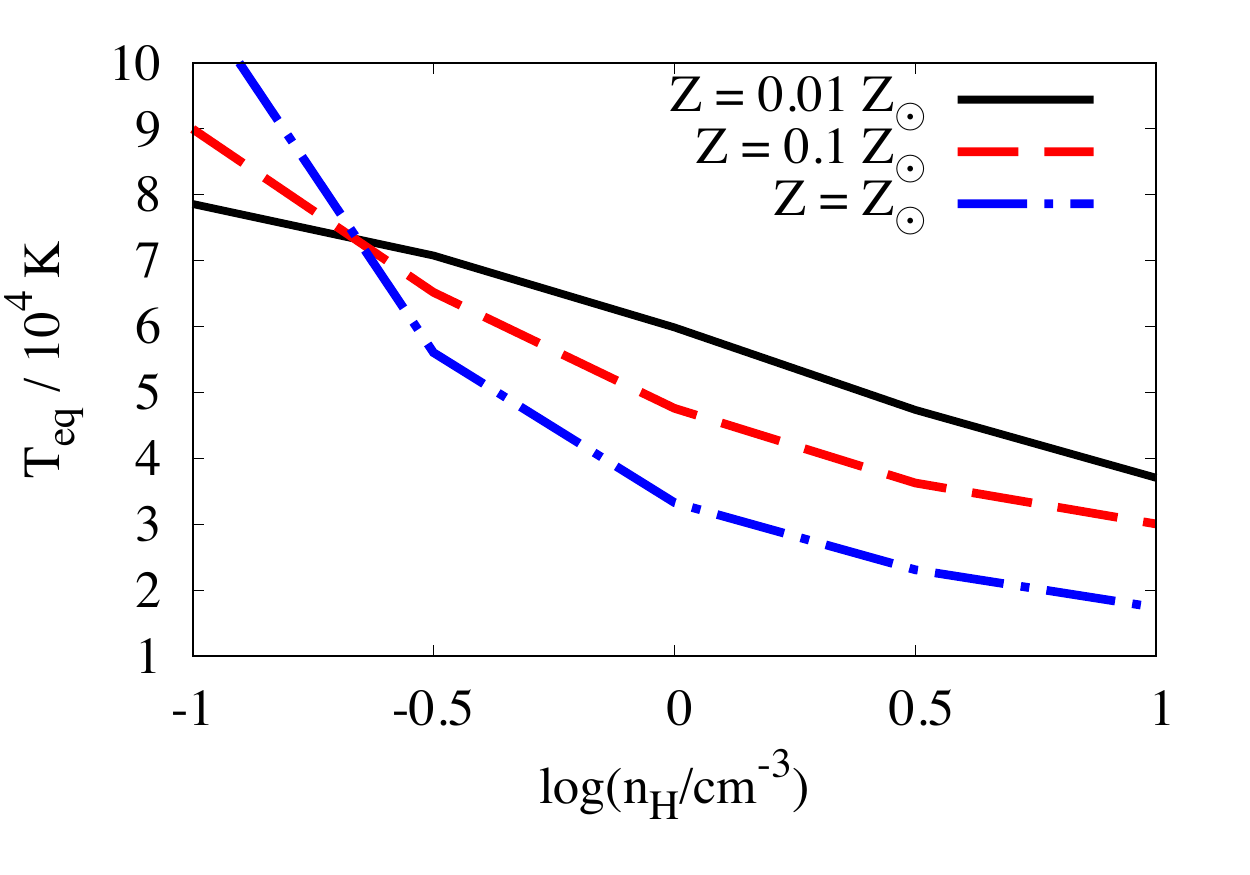}
\caption{Equilibrium temperature of photo-ionized CGM as a function of density and metallicity, assuming a `standard' QSO spectrum with the median luminosity of the QSOs of \citealt{Borisova+16} and a distance from the source $d = 50 \; \textrm{kpc}$. An equilibrium temperature as low as $2 \times 10^4 \; \textrm{K}$ can be achieved, in close proximity to the QSO, only at high densities and solar metallicity. \label{fig::Teq}}
\end{figure}

Deviations from the simple estimate above would mostly come from (i) line cooling from residual non-completely ionized hydrogen and helium and (ii) traces of metals, which would constitute additional line cooling channels, but also additional targets to photo-electric heating. Other potentially relevant effects are those whereby the energy of a photon can be distributed to more than one electron and primarily (i) ionization of hydrogen by high-energy HeII recombination photons and (ii) secondary ionization (of hydrogen or other species) by high-energy electrons produced in a previous ionization event (\citealt{OsterbrockFerland}). To assess the role of all these effects, we performed more detailed calculations of the equilibrium temperature using the photo-ionization code Cloudy (version 13.04, last described by \citealt{Ferland2013}). We modelled the hydrogen and helium ionizing spectrum as a single power law with slope $p = 1.7$ . We also included a hardening at X-ray frequencies (constant $\nu L_\nu$ in the energy range $2 \; \textrm{keV} < E < 100 \; \textrm{keV}$) and a cut-off for $ E > 100 \; \textrm{keV}$, though we found the details of the X-ray spectrum to be unimportant. We adopted a specific luminosity at the Lyman limit  $\tilde{L}_{\nu, \textrm{LL}} = 5.2 \times 10^{31} \; \textrm{erg} \; \textrm{s}^{-1} \; \textrm{Hz}^{-1}$ (see Section \ref{subsec::H} below) and a distance from the source $d = 50 \; \textrm{kpc}$. Results are shown in Figure \ref{fig::Teq} for different values of gas density and metallicity. 

We see that the results are in broad agreement with the simple analytical estimate above (equation \ref{Teq}) especially for intermediate densities and metallicities. As also expected, the equilibrium temperature decreases with increasing density, due to collisionally excited line cooling, the effect being larger for larger metallicity. Note that at low densities the effect of non-primordial metallicity is reversed, as photo-electric heating on highly ionized metal species (as well as a small contribution of Compton heating) dominates over additional line cooling, so that the equilibrium temperature is even larger than the analytic estimate.

An equilibrium temperature equal or smaller than $T \sim 2 \times 10^4 \; \textrm{K}$, as desirable from inspection of Figure \ref{fig::Tplot}, can be reached only for a rather high density $n = 10 \; \textrm{cm}^{-3}$ and solar metallicity. Note that at a distance $d = 50 \; \textrm{kpc}$ and a temperature $T = 2 \times 10^4 \; $, a density $n = 10 \; \textrm{cm}^{-3}$ corresponds to a filling factor $f_V = 7.3 \times 10^{-6}$ (equation \ref{densprofile}), which would give an absurdly large baryon fraction even at low temperature (Figure \ref{fig::Tplot}). Furthermore, it is not obvious that a metallicity as high as solar can be easily reached in the CGM at $z = 3$. Doing the calculation at smaller distances $d < 50 \; \textrm{kpc}$ would not help in obtaining lower temperatures, as the ionization parameter increases with decreasing radius for our density slope $\gamma = 1.25 < 2$. 

As a sanity check, we also repeated our calculation including only the (QSO-dominated) ionizing background at $z = 3.2$, a metallicity $Z = 0.1 \; Z_\odot$ and a density equal to 10 times the critical density at $z = 3.2$, in order to mimic the conditions in the IGM far away from an individual QSO. We found in this case an equilibrium temperature $T_\textrm{eq} = 3 \times 10^4 \; \textrm{K}$, in agreement with models of HeII re-ionization (e.g.\ \citealt{McQuinn+09}; \citealt{UptonSanderbeck+16}). The relatively low equilibrium temperature in this case is due to the relatively low ionization parameter at large distances from a single bright QSO. This emphasizes that temperatures as high as those shown in Figure \ref{fig::Teq} are expected, for a gas photo-ionized by a QSO, only in the very close vicinity to the source, such as in the CGM of the host galaxy, which is the main focus of this work.

\begin{figure}
\centering
\includegraphics[width=9cm]{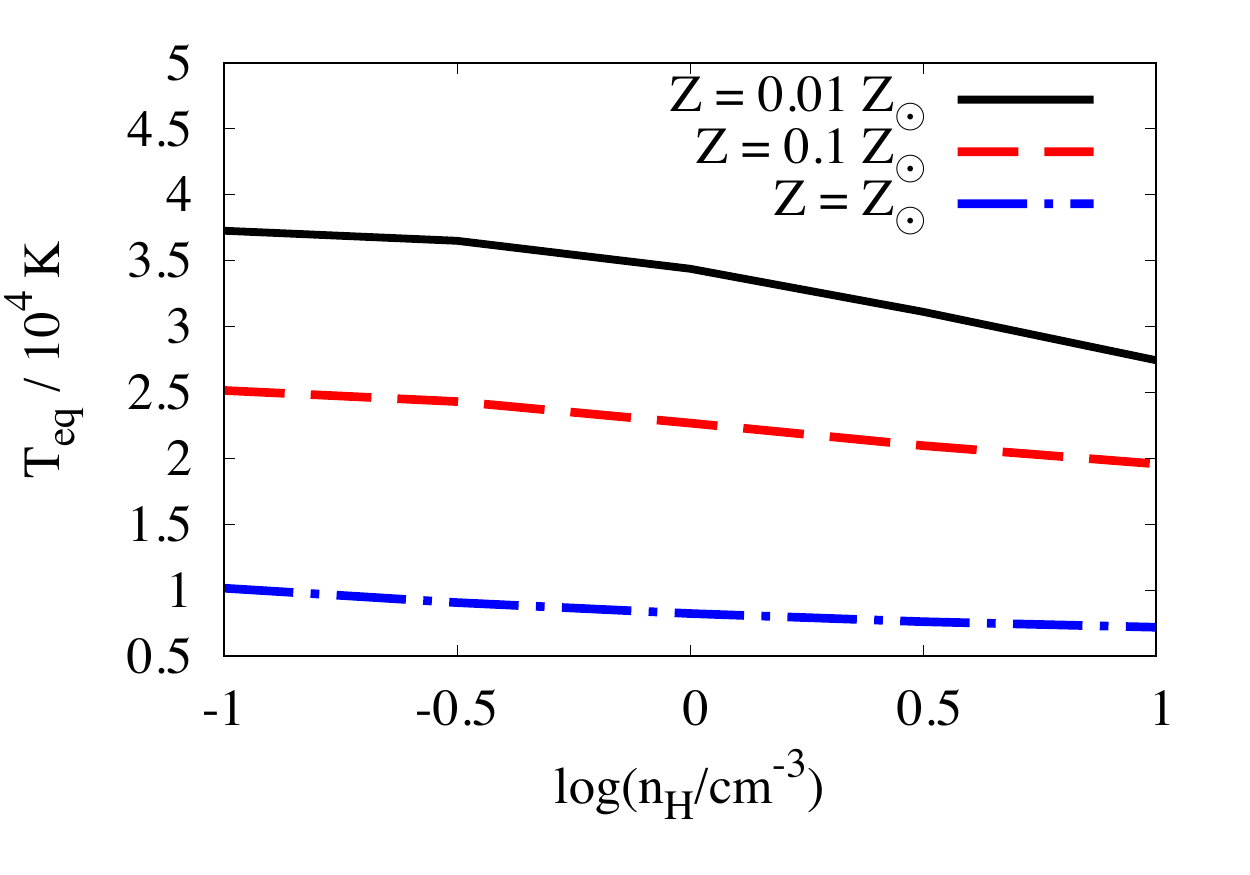}
\caption{Similar to Figure \ref{fig::Teq}, but assuming that the incident radiation field is exclusively due to the thermal radiation of a classical (geometrically thin, optically thick) accretion disc, accreting at the Eddington rate around a black hole with mass  $M_{BH} = 10^8 \; \textrm{M}_\odot$. Note that this model neglects any comptonization effect, with resultant strong suppression of EUV and X-ray photons. \label{fig::Teq_puredisc}}
\end{figure}

Another possibility to achieve lower equilibrium temperatures at the small distances of interest could be to change the shape of the ionizing spectrum, especially at energies $E > 4 \; \textrm{Ry}$, which contribute most of the heating and on which observational constraints are very scant. We therefore also considered a modified spectrum, identical to the one above but for being double as steep at $E > 4 \; \textrm{Ry}$ than in the range $1 \; \textrm{Ry} < E < 4 \; \textrm{Ry}$. However, both analytic estimate and Cloudy modelling showed in this case only a modest $\sim 25 \%$ decrease in $T_\textrm{eq}$. Virtually, only an exponential cut-off in the EUV spectrum would be effective in beating the temperature down as required. We note that this is in fact the theoretical expectation for the emission spectrum of a classical accretion disc around a supermassive black hole (see e.g.\ \citealt{ShakuraSunyaev1973}; \citealt{LaorDavis2011}). The predictions of this kind of model are illustrated in Figure \ref{fig::Teq_puredisc}, where we assumed the spectrum of a classical (geometrically thin, optically thick) accretion disc (\citealt{ShakuraSunyaev1973}) with Eddington accretion around a supermassive black hole with mass $M_{BH} = 10^8 \; \textrm{M}_\odot$, ignoring in particular any comptonization effect. As expected, this model easily predicts an equilibrium temperature $T_\textrm{eq} \simeq 2 \times 10^4 \; \textrm{K}$ for a metallicity $Z = 0.1 \; Z_\odot$ and a broad range of densities and even lower temperatures ($T_\textrm{eq} \simeq 10^4 \; \textrm{K}$) for solar metallicity. This kind of solution is formally appealing and it emphasizes the crucial role of assumptions on the shape of the ionizing spectrum of QSOs. Nonetheless, we refrain from considering this option as our fiducial case, as it would raise significant issues with a large body of independent observations including (i) widespread evidence of comptonization in AGN (e.g.\ \citealt{Done+12} and references therein) and (ii) indirect evidence based on HeII re-ionization (e.g.\ \citealt{Worseck+16} and references therein).

Motivated by the discussion above, in the following we will sometimes adopt as fiducial the value $T_\textrm{cold} = 5.6 \times 10^4 \; \textrm{K}$ (cf.\ equation \ref{Teq} and Figure \ref{fig::Teq}). We will nonetheless also present our results for a large range of temperatures to allow for more generality.

\subsection{The minimum CGM baryon fraction}\label{subsec::minfbtot}

\begin{figure}
\centering
\includegraphics[width=9cm]{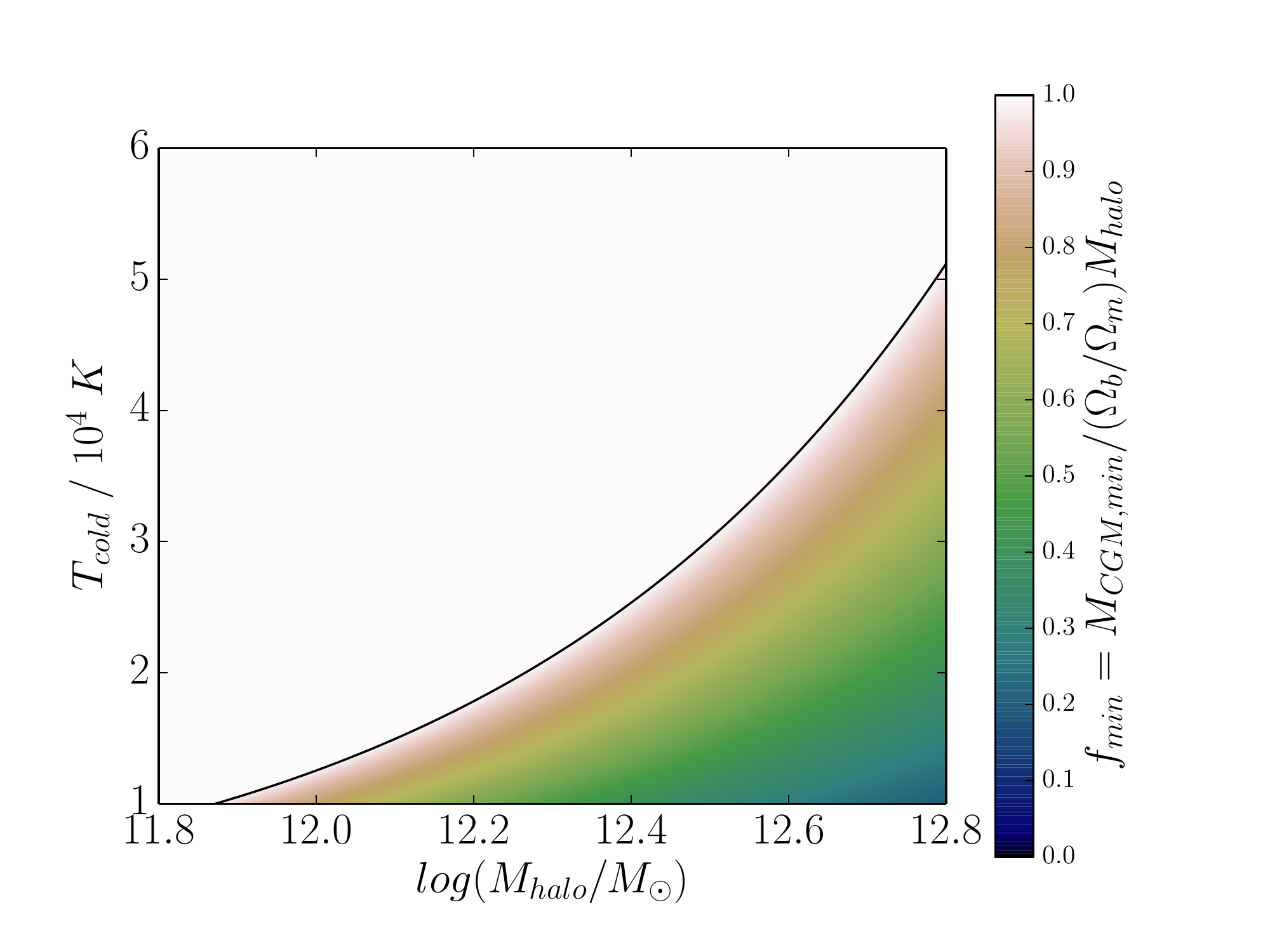}
\caption{\emph{Minimum} CGM baryon fraction as a function of the assumed halo mass and temperature of the cold gas. Models with low halo masses and high temperatures (blank) are unphysical, as they require a baryon fraction larger than cosmological. Conversely, a CGM baryon fraction significantly smaller than cosmological requires both high halo masses and low gas temperatures, possibly in tension with constraints from QSO clustering and photo-heating (Sections \ref{subsec::HaloMass} and \ref{subsec::Tcold}). \label{fig::minplot}}
\end{figure}

A useful way to summarize the results of the previous sections is to compute, for each value of the parameters $M_\textrm{halo}$ and $T_\textrm{cold}$, the \emph{minimum} total CGM baryon fraction. As it is readily found from equation \eqref{mutot}, this is equal to
\begin{equation}\label{mumin}
f_\textrm{min} = 2 f_1 \varepsilon_1^{1/2} T_4^{(1+t)/2} M_{12}^{-(1+\gamma)/3}
\end{equation}
and is achieved for the filling factor $f_{V,\textrm{eq}}$ such that cold gas and hot gas contribute equally to the total mass:
\begin{equation}
f_{V, \textrm{eq}} = \varepsilon_1 \; T_4 M_{12}^{-2/3} \; .
\end{equation}
Equation \eqref{mumin} gives the \emph{lower limit} to the total CGM baryon fraction needed to explain the observations within our assumptions. It is shown in Figure \ref{fig::minplot}, as a function of the two parameters $M_\textrm{halo}$ and $T_\textrm{cold}$. The solid black line indicates the cosmological baryon fraction. The blank region at low masses and large temperatures corresponds to unphysical solutions (baryon fraction larger than cosmological), while progressively smaller baryon fractions are inferred at large halo masses and low temperatures. 

The couple of fiducial values $M_\textrm{halo} \simeq 10^{12} \; \textrm{M}_\odot$ and $T_\textrm{cold} \simeq 5.6 \times 10^4 \; \textrm{K}$ (determined in Sections \ref{subsec::HaloMass} and \ref{subsec::Tcold}, respectively) fall in the unphysical region. The simplest explanation for this fact is that at least one of these values is incorrect.  We do not have strong reasons to prefer a revision of either $M_\textrm{halo}$ or $T_\textrm{cold}$. We stress, however, that both parameters can be constrained by independent observations and in particular by further studies of clustering and spectral properties of luminous QSOs (Sections \ref{subsec::HaloMass} and \ref{subsec::Tcold}). Another option is that both our fiducial values are correct, but that some of our other assumptions (recombination-dominated emission and pressure equilibrium) is wrong. These possibilities are discussed in Sections \ref{sec::Scattering} and \ref{sec::Discussion}, but we anticipate that they are not trivially achieved either.

Figure \ref{fig::minplot} can also be compared to cosmological hydrodynamical simulations or to semi-analytic models, as long as these include predictions for the CGM baryon fraction in massive haloes at high redshift.

Cosmological hydrodynamical simulations with strong ejective feedback (see Section \ref{sec::Introduction}) typically predict a CGM baryon fraction equal to $\sim 30-40 \%$ of the cosmic average at the relevant halo mass and redshift (e.g.\ \citealt{Dave2009}; \citealt{Liang+16}; \citealt{Kulier+19}; Mitchell, private communication). Unfortunately, not all simulation projects have published suitable predictions, so we cannot make an exhaustive comparison here. On the other hand, we hope that our results can encourage more of these predictions in the future. Based on the references above and on Figure \ref{fig::minplot}, we expect that simulations with strong ejective feedback should be able to reproduce the observed Ly$\alpha$ surface brightness, under the assumption of fluorescent emission (recombination), only if the analysis is performed on rather massive haloes ($M_\textrm{halo} \gtrsim 10^{12.5} \; \textrm{M}_\odot$) \emph{and} the cold CGM temperature is assumed to be sufficiently low ($T_\textrm{cold} \lesssim 2 \times 10^4 \; \textrm{K}$), despite photo-heating by the central QSO. Furthermore, it is required that the CGM mass is split almost equally among the cold phase and the hot phase, so that the former is both massive and dense enough to give rise to sufficiently bright recombination radiation. In Section \ref{subsec::Gronke} we discuss in some detail the only experiment published so far that can be directly compared to our expectations.

Among semi-analytic models with strong ejective feedback, the one of \cite{Henriques+19} predicts that a typical halo with mass $M_\textrm{halo} = 10^{12} \; \textrm{M}_\odot$ at $z = 3$ would have a median of 53\% of its baryons in the CGM. This is slightly larger than what found by the simulations mentioned above and therefore potentially closer to what is needed to explain the observations. Note however that the CGM in the model of \cite{Henriques+19} is all hot and therefore invisible in Ly$\alpha$, unless a large fraction of it is somehow assumed to be converted into cold CGM.

\subsection{Comparison with \citealt{GronkeBird2017}}\label{subsec::Gronke}
A special mention is needed to the study of \cite{GronkeBird2017}. These authors analysed massive haloes at $z = 2$ in the Illustris simulation (\citealt{Vogelsberger+14}) and calculated the expected recombination Ly$\alpha$ surface brightness, under the assumption that the CGM is highly ionized. They found agreement with the observed surface brightness profile of the Slug nebula and of MUSE Giant Ly$\alpha$ nebulae (note that the \citealt{Borisova+16} profile in their Figure 9 should be shifted upwards by 0.58 dex due to differential redshift-dimming between the observed $z = 3.2$ and the simulated $z = 2$). As they assumed recombination radiation from a highly ionized medium, their experiment is directly comparable to our analytic calculations. As discussed in Section \ref{subsec::minfbtot}, we expect simulations with strong ejective feedback to be able to reproduce a Ly$\alpha$ surface brightness as high as observed, as long as the assumed halo mass and gas temperature are sufficiently large and small, respectively. Indeed, \cite{GronkeBird2017} adopted a range of halo masses centred on $M_\textrm{halo} = 10^{12.5} \; \textrm{M}_\odot$, which is on the high side of those discussed in Section \ref{subsec::HaloMass}. Furthermore, their post-processing analysis neglected photo-heating from the QSO, with the result that they most likely underestimated the temperature of the photo-ionized CGM. Note that Illustris does include a recipe for radiative feedback from QSOs (\citealt{Vogelsberger+13}). \cite{GronkeBird2017}, however, did not select their haloes on the requirement that the central super massive black hole be in an active phase and therefore a QSO would be found therein with a probability that can be approximated by the QSO duty cycle $f_\textrm{duty} \ll 1$ (e.g.\ \citealt{HaimanCiottiOstriker2004}; \citealt{Eftekharzadeh+15}). Note also that any imprint of past QSO activity on photo-ionized gas temperatures would die out on roughly one recombination time-scale, or less than $1 \; \textrm{Myr}$ in our case.

We therefore think that the formal success found by this work is not surprising and actually in line with our analytic expectations, but it may be driven by a particularly favourable choice of the parameters. In particular, our analysis shows that neglecting in post-processing photo-heating of the cold CGM by the central QSO can significantly bias the recombination rate and therefore the predicted Ly$\alpha$ surface brightness in the direction needed to make the comparison with observations easier.

\cite{GronkeBird2017} also considered the option that most of the CGM is ionized by star formation rather than by the central QSO, in which case the temperature of the photo-ionized gas can be lower than our estimates. This option, however, is strongly disfavoured for the MUSE Giant Ly$\alpha$ nebulae, as we have argued already in Section \ref{subsubsec::ionstars}.

\section{Ionization structure}\label{sec::ion}

Here we verify the self-consistency of our model. We start from the key requirement that the CGM can be almost completely photo-ionized by the central QSO (Section \ref{subsec::H}), before discussing the ionization of helium (Section \ref{subsec::He}) and the possible contribution of collisional excitation to the Ly$\alpha$ emission (Section \ref{subsec::collex}). We obviously refer here only to the fraction of the solid angle that is illuminated by the QSO. We recall that considering any substantial QSO obscuration (by a dusty torus or any other intervening material) would only go in the direction of reinforcing our conclusions (see Section \ref{subsec::caveats}).

\subsection{Hydrogen}\label{subsec::H}
The QSOs in the sample of \cite{Borisova+16} have a median specific luminosity at the Lyman limit frequency $\nu_\textrm{LL}$ of $\tilde{L}_{\nu, \textrm{LL}} = 5.2 \times 10^{31} \; \textrm{erg} \; \textrm{s}^{-1} \; \textrm{Hz}^{-1}$, which we estimated following \cite{FAB15} and using the absolute i-band magnitudes of the QSOs in the \cite{Borisova+16} sample (see their Table 2). The corresponding hydrogen ionization rate at distance $r$ from the QSO is
\begin{equation}\label{Gamma}
\left. \begin{array}{ll}
\Gamma_\textrm{ion} (r) & = \dfrac{1}{4 \pi r^2} \displaystyle\int_{\nu_\textrm{LL}}^{+\infty} \dfrac{L_\nu}{h \nu} \sigma^\textrm{ion}(\nu) \; d \nu \\
&\\
& = \dfrac{1}{p+3} \dfrac{L_{\nu, \textrm{LL}} \; \sigma^\textrm{ion}_\textrm{LL}}{4 \pi h r^2} \\
&\\
& = 8.8 \times 10^{-7} \xi \left( \dfrac{r}{10 \; \textrm{kpc}} \right)^{-2} \; \textrm{s}^{-1} \; ,
\end{array}\right.
\end{equation}
where $\sigma^\textrm{ion}(\nu) \simeq \sigma^\textrm{ion}_\textrm{LL} \; (\nu/\nu_\textrm{LL})^{-3} $, with  $\sigma^\textrm{ion}_\textrm{LL} = 6.3 \times 10^{-18} \; \textrm{cm}^2$, is the hydrogen photo-ionization cross section and $L_\nu \simeq  L_{\nu, \textrm{LL}}  (\nu/\nu_\textrm{LL})^{-p}$ is a power-law approximation of the ionizing continuum of the QSO, with a typical slope $p = 1.7$ (e.g. \citealt{Lusso+15}), while the factor
\begin{equation}\label{csidef}
\xi \equiv  \left( \frac{L_{\nu, \textrm{LL}}}{5.2 \times 10^{31} \; \textrm{erg} \; \textrm{s}^{-1} \; \textrm{Hz}^{-1}} \right) \left( \frac{p+3}{4.7}  \right)^{-1} \; ,
\end{equation}
incorporates possible deviations of the normalization and slope of the QSO spectrum from our fiducial values. Note that adopting a more precise double power-law approximation to the frequency dependence of the hydrogen ionization cross section (see e.g.\ \citealt{Meiksin2009}) is equivalent to replacing in the expressions above $\frac{1}{p+3} \rightarrow \frac{p + s + \beta}{(p + s)(p+s+1)}$, with $s = 2.99$ and $\beta = 1.34$. This gives a correction of less than $7 \%$ for any $p > 1$ and is negligible to our purposes.

The photo-ionization rate given in equation \eqref{Gamma} should be compared to the (case A) \emph{total} hydrogen recombination rate (per proton) for a fully ionized gas:
\begin{equation}\label{Gammarec}
\left.\begin{array}{ll}
\Gamma_\textrm{rec} & = \alpha_\textrm{rec}(T) \left( 1 + 2\dfrac{n_{He}}{n_H} \right) n \\
& \\
& = 4.8 \times 10^{-13} \; T_4^{-0.75} \left( \dfrac{n}{\textrm{cm}^{-3}} \right) \; \textrm{s}^{-1} \; ,
\end{array}\right.
\end{equation}
where again we adopted a power-law approximation to a more complex temperature dependence (\citealt{Meiksin2009}) over the temperature range of interest and we accounted for electrons from doubly ionized helium. It is readily seen that $\Gamma_\textrm{rec} \ll \Gamma_\textrm{ion}$ for any plausible density for CGM conditions (the equality would require $n \simeq 2 \times 10^4 \; \textrm{cm}^{-3}$ at $r = 100 \; \textrm{kpc}$ and even larger densities at smaller radii). Note also that $\Gamma_\textrm{ion}$ exceeds the photo-ionization rate  due to the cosmic UV background (e.g.\ \citealt{HM12}) by several orders of magnitude at any distance plausibly associated with the CGM.

Again following \cite{Meiksin2009}, the equilibrium hydrogen neutral fraction under these conditions is achieved on very short time-scale $t_\textrm{ion} = \Gamma_\textrm{ion}^{-1}$ (everywhere smaller than a few years in our case) and very well approximated by $x_{HI} = \Gamma_\textrm{rec} / \Gamma_\textrm{ion}$, which gives
\begin{equation}\label{xHI}
x_{HI} = 7.9 \times 10^{-8} f_V^{-1/2}  \xi^{-1} T_4^{-0.27} \left( \dfrac{r}{10 \; \textrm{kpc}} \right)^{0.75} \; .
\end{equation}
Combining equations \eqref{densprofile} and \eqref{xHI}, we obtain the neutral hydrogen number density $n_{HI} = x_{HI}n$, which reads
\begin{equation}\label{avenHIprofile}
\left.\begin{array}{l}
n_{HI} = 1.1 \times 10^{-8} \; f_V^{-1} \; \xi^{-1} \; T_4^{0.21} \left( \dfrac{r}{10 \; \textrm{kpc}} \right)^{-0.5} \; \textrm{cm}^{-3} \; ,
\end{array}\right.
\end{equation}
and therefore the \emph{volume-averaged} neutral hydrogen number density
\begin{equation}
\langle n_{HI} \rangle = f_V n  = 1.1 \times 10^{-8} \; \xi^{-1} \; T_4^{0.21} \left( \dfrac{r}{10 \; \textrm{kpc}} \right)^{-0.5} \; \textrm{cm}^{-3} \; ,
\end{equation}
which is \emph{independent} on the filling factor. This is not surprising, if one considers that both recombination radiation (on which our inferred density profile of equation \ref{densprofile} is based) and the neutral fraction are determined by the same physical process (obviously, recombination). It is similarly easy to show that the inferred \emph{volume-averaged neutral} hydrogen number density $\langle n_{HI} \rangle$ is actually independent also on assumptions on the ionization state of helium (which enters in both equations \ref{Bdef} and \ref{Gammarec}) and on possible internal clumpiness $C_\textrm{int}$ within the cold gas structures (cf.\ Section \ref{subsubsec::fV}), as all these factors equally contribute to the recombination emissivity and the recombination rate. The small dependence on temperature is due to a non-constant efficiency of Ly$\alpha$ photons production per recombination. The main dependence is, of course, on the intensity and spectrum of the ionizing source, as parametrized here by the factor $\xi$ (equation \ref{csidef}).

If radially integrated, equation \eqref{avenHIprofile} gives the neutral hydrogen column density between the QSO and a point in the nebula at distance $r$ from the centre:
\begin{equation}\label{NHI}
N_\textrm{HI} (r) = 7.1 \times 10^{14} \xi^{-1} T_4^{0.21} \left( \dfrac{r}{10 \; \textrm{kpc}} \right)^{0.5} \; \textrm{cm}^{-2} \; ,
\end{equation}
which is everywhere much smaller than the Lyman limit column density $N_{H, LL} = \sigma_\textrm{ion}(\nu_{LL})^{-1} = 10^{17.2} \; \textrm{cm}^{-2}$, for any plausible value of $\xi$ and $T_4$. This confirms that our main assumption that the entire nebula is almost completely ionized and transparent to hydrogen ionizing radiation is self-consistent. The implications for Ly$\alpha$ scattering, on the other hand, are more complicated and are addressed in detail in Section \ref{sec::Scattering}.

\subsection{Helium}\label{subsec::He}
Similarly, the column density of HeII is
\begin{equation}\label{NHeII}
N_{HeII} = 1.3 \times 10^{16} \xi_{He}^{-1} T_4^{0.27} \left( \dfrac{r}{10 \; \textrm{kpc}} \right)^{0.5} \; \textrm{cm}^{-2} \; ,
\end{equation}
with
\begin{equation}
\xi_{He} \equiv 4^{-(p_1-1.7)} \; \frac{p_1+3}{p_2+3} \; \xi \; ,
\end{equation}
where we used again atomic coefficients as in \cite{Meiksin2009} and we further assumed that the filling factor $f_V > 10^{-6}$, while $p_1$ and $p_2$ are the slopes of the QSO ionizing continuum in the ranges $1 \; \textrm{Ry} < E < 4 \; \textrm{Ry}$ and $E > 4 \; \textrm{Ry}$, respectively. If for simplicity we assume $p_2 = p_1 = 1.7$ (i.e.\ if the spectrum by \citealt{Lusso+15} can be extrapolated above 4 Ry), we see that the column density \eqref{NHeII} is still safely below the limit for self-shielding $N_{HeII, \textrm{crit}} = 6.3 \times 10^{17} \; \textrm{cm}^{-2}$, implying that the entire nebula should be transparent to the helium-ionizing radiation as well. 

Note, however, that \cite{Borisova+16} estimated very low HeII/Ly$\alpha$ line ratios, inconsistent with recombination radiation from fully ionized hydrogen and helium. Though the estimate above suffers from several approximations and extrapolations (most notably, the shape of the ionizing spectrum and the extrapolation of the modelled density and filling factor to radii smaller than 10 kpc), we found that it is not easy to accommodate HeII column density that are high enough to guarantee self-shielding. If this is true, and if the very low limits on the line ratios are confirmed by deeper observations, this could suggest that either some of the Ly$\alpha$ emission is contributed by processes other than pure recombination (see Sections \ref{subsec::collex} and \ref{sec::Scattering} below; see also \citealt{Humphrey+19}), or that the majority of the emission comes from regions that are dense enough that the helium is only partially ionized despite being fully illuminated (e.g.\ \citealt{FAB15}). This in our case would imply a very low filling factor $f_V < 10^{-6}$, or additional internal clumpiness within the cold gas structures (see Section \ref{subsec::overP} below; see also \citealt{Cantalupo+19}). It is of course also possible that the explanation is a combination of those mentioned above and it may be different for different nebulae. We plan to explore these aspects in more detail in future work.

\subsection{Collisional excitation}\label{subsec::collex}

We finally verify that the contribution of collisional excitation to the Ly$\alpha$ emission can be consistently neglected in our model.

The local Ly$\alpha$ emissivity due to collisional excitation and recombination are, respectively:
\begin{equation}
j_\textrm{coll} = \frac{E_{\textrm{Ly}\alpha}}{4 \pi} \alpha^\textrm{eff}_\textrm{coll} n_\textrm{HI} n_e
\end{equation}
and
\begin{equation}
j_\textrm{rec} = \frac{E_{\textrm{Ly}\alpha}}{4 \pi} \alpha^\textrm{eff}_\textrm{rec} n_\textrm{HII} n_e
\end{equation}
(we omitted `Ly$\alpha$' super-scripts or sub-scripts j's and $\alpha$'s for clarity) and their ratio is
\begin{equation}
\frac{j_\textrm{coll}}{j_\textrm{rec}} = \frac{\alpha^\textrm{eff}_\textrm{coll}}{\alpha^\textrm{eff}_\textrm{rec}} \frac{x_{HI}}{1 - x_{HI}} \; ,
\end{equation}
where we take $x_HI$ from equation \eqref{xHI} and the ratio of recombination to collisional excitation coefficients as in \cite{Cantalupo+08}. The ratio $\alpha^\textrm{eff}_\textrm{coll} / \alpha^\textrm{eff}_\textrm{rec}$ increases rapidly with temperature (much more rapidly than the decrease with temperature of $x_{HI}$, see equation \ref{xHI}). To obtain a conservative upper limit to the contribution of collisional excitation, we therefore consider a fairly high temperature $T = 5 \times 10^4 \; \textrm{K}$ and also a fairly large distance from the centre $r = 100 \; \textrm{kpc}$ (to maximize the neutral fraction). Note that choosing even higher temperatures would not help, as collisional ionization would start to dominate over photo-ionization driving an additional drop of the neutral fraction and therefore of the collisional excitation rate. In these conservative conditions, we obtain as an upper limit:
\begin{equation}
\frac{j_\textrm{coll}}{j_\textrm{rec}} < 1.4 \times 10^{-2} \xi^{-1} f_V^{-1/2}
\end{equation}
so that we can safely ignore collisional excitation as a powering mechanism for filling factors larger than $10^{-4}$ (a limit that goes further down for smaller distances and temperatures).

\section{The impact of scattering}\label{sec::Scattering}

Here we discuss the possible impact on our results of Ly$\alpha$ scattering. We distinguish between the \emph{addition} of Ly$\alpha$ photons, which may be produced by the central QSO and scattered into the line of sight (Section \ref{subsec::BLR}), from the \emph{redistribution} of Ly$\alpha$ photons due to scattering within the CGM (Sections \ref{subsec::InternalScattering} and \ref{subsec::NeutralScreen}).

\subsection{Scattering from the broad-line region}\label{subsec::BLR}

Even if the CGM is almost completely ionized, some of the very rare neutral hydrogen atoms can still intercept some of the Ly$\alpha$ photons coming from the broad-line region (BLR) of the central QSO and scatter them in to the line of sight. These photons would contribute to the observed Ly$\alpha$ surface brightness of the nebulae, in addition to the recombination radiation we have focused on so far.

To quantify this effect, we must convert the neutral hydrogen column density \emph{as seen by the QSO} (equation \ref{NHI}) into a line-centre optical depth to Ly$\alpha$ scattering (following e.g.\ \citealt{Meiksin2009}). We find
\begin{equation}\label{tau}
\tau_{\textrm{Ly}\alpha, \textrm{QSO}} (r) = 0.99 \; \xi^{-1} T_4^{0.21} \sigma_{300}^{-1} \left( \frac{r}{10 \; \textrm{kpc}} \right)^{0.5} \; ,
\end{equation}
where $\sigma_{300}$ is the 1D velocity dispersion of the cold CGM, \emph{as seen by the QSO}, in units of 300 $\textrm{km} \; \textrm{s}^{-1}$. 

We have therefore found a model-dependent optical depth of order unity. This is the most difficult regime for analytic treatment, as both the single scattering and the diffusion limit are prone to be inadequate. On the other hand, a brute-force numerical approach would not give a clear-cut solution either, as the limiting factor here is not the calculation itself, but rather that the result is bound to be strongly dependent on the \emph{assumptions} on the intrinsic kinematics of the cold gas, which is largely unknown.

The reference value for $\sigma$ that we used in equation \eqref{tau} corresponds to the typical width of the Ly$\alpha$ line as observed in emission (\citealt{Borisova+16}). Note however that this can be a biased estimate of the true value of $\sigma$ for at least two reasons. First, if radiative transfer effects were indeed important (which is what we would like to establish here), the observed line width would likely be larger than the intrinsic one, with the result that the fiducial $\sigma$ above would be overestimated and the optical depth underestimated. Second, the velocity dispersion \emph{as seen by the QSO} will be different from that \emph{along the line of sight} unless the kinematics of the cold CGM is isotropic. For instance, if the cold CGM were dominated by coherent radial flows (such as inflows or outflows) the dispersion seen by the QSO (as determined by the acceleration or deceleration of the cold gas flows) would again be smaller than that along the line of sight (which would instead be dominated by projection effects). Both effects would go in the direction of favouring optical depths larger than unity. As we show below, a large optical depth would imply a small contribution of BLR photons to the observed emission and therefore that the model considered so far is self-consistent in this respect (although not necessarily unique).

Our ignorance on the intrinsic kinematics of the CGM prevents us from calculating the true contribution of scattered BLR photons to the observed emission. We can none the less easily put a very conservative upper limit to this contribution by calculating the \emph{maximum possible} number of BLR Ly$\alpha$ scattered photons per fluorescent Ly$\alpha$ photon produced in a recombination event, based on fundamental atomic physics, as described below (we have also verified the validity of our reasoning  with Cloudy simulations).

Recombination Ly$\alpha$ photons are produced, with some probability $\eta$, every time an electron recombines with a proton to form an atom of neutral hydrogen. Statistically, the electron will be freed again (and therefore become again eligible to recombination and Ly$\alpha$ emission) in a time equal to the inverse of the photo-ionization rate $t_\textrm{ion} = \Gamma_\textrm{ion}^{-1}$. During this time, the neutral atom is available to undergo a number of scattering events, statistically equal to $\Gamma_\textrm{scatt} t_\textrm{ion} = \Gamma_\textrm{scatt}/\Gamma_\textrm{ion}$, where $\Gamma_\textrm{scatt}$ is the scattering probability per unit time
\begin{equation}\label{Gammascatt}
\Gamma_\textrm{scatt} = \int_0^{+\infty} \frac{F_\nu (\nu)}{h \nu} \sigma_\textrm{scatt} (\nu) d \nu = \frac{F_{\nu, \textrm{Ly}\alpha}}{h\nu_{\textrm{Ly}\alpha}} \bar{\sigma} \; ,
\end{equation}
where $\bar{\sigma} = \int_0^{+\infty} \sigma_\textrm{scatt} (\nu) d \nu = 1.1 \times 10^{-2} \; \textrm{cm}^2 \; \textrm{Hz}$ is the frequency-integrated cross-section to Ly$\alpha$ scattering (e.g.\ \citealt{RybickiLightman}; \citealt{Meiksin2009}) and
\begin{equation}\label{attenuatedF}
F_{\nu, \textrm{Ly}\alpha} = \frac{L_{\nu, \textrm{Ly}\alpha}}{4 \pi r^2} e^{-\tau_{\textrm{Ly}\alpha, \textrm{QSO}} (r)}
\end{equation}
is the specific flux density of Ly$\alpha$ photons propagating \emph{directly} from the BLR to distance $r$ from the QSO. The cumulative Ly$\alpha$ optical depth $\tau_{\textrm{Ly}\alpha, \textrm{QSO}} (r)$ describes attenuation by scattering at smaller radii.\footnote{Obviously attenuation here is not synonym to absorption. The fate of intercepted BLR photons after first scattering is addressed in Section \ref{subsec::InternalScattering}.} Note that we do not need to introduce an attenuation factor for the Lyman limit photons, as the optical depth to photo-ionization is negligible irrespective of assumptions on kinematics, at the column densities of interest (Section \ref{subsec::H}). 

The local ratio of the scattering to recombination Ly$\alpha$ emissivity is given by
\begin{equation}\label{jratio}
\frac{j_\textrm{scatt}}{j_\textrm{rec}} = \frac{\Gamma_\textrm{scatt}}{\eta \Gamma_\textrm{ion}} = \frac{p+3}{\eta} \frac{\bar{\sigma}}{\sigma^\textrm{ion}_{LL} \nu_{\textrm{Ly}\alpha}} \frac{F_{\nu, \textrm{Ly}\alpha}}{F_{\nu, \textrm{LL}}} = 0.71 \frac{p+3}{\eta} \frac{F_{\nu, \textrm{Ly}\alpha}}{F_{\nu, \textrm{LL}}} \; ,
\end{equation}
where the case A Ly$\alpha$ production efficiency per recombination is well approximated by
\begin{equation}
\eta(T) = 0.41 \; T_4^{-0.23} \; .
\end{equation}
It is clear from equations \eqref{attenuatedF} and \eqref{jratio} that a very conservative upper limit for the ratio of scattering to recombination emissivity is found in the Ly$\alpha$ optically thin limit $\tau_{\textrm{Ly}\alpha, \textrm{QSO}} \ll 1$. Furthermore, in this limit, the ratio $j_\textrm{scatt}/j_\textrm{rec}$ is just a constant, determined by the intrinsic spectrum of the QSO. This is not surprising if one considers that, in an almost completely ionized gas, one single physical mechanism (recombination) is responsible both for Ly$\alpha$ fluorescence and for the creation of scattering targets (neutral atoms) and that in both cases the energy source is the same (the QSO). Assuming the typical QSO spectrum from \cite{Lusso+15} and a UV slope $p = 1.7$ we then find
\begin{equation}\label{Rdef}
\mathcal{R} \equiv \frac{j_\textrm{scatt}}{j_\textrm{rec}} \leq \mathcal{R}_\textrm{max} (T) = 26 \; T_4^{0.23} \; .
\end{equation}
We stress that the value above is a very strict upper limit on the true ratio of the scattering-to-recombination emissivity, as it assumes that the whole CGM around the QSO sees the Ly$\alpha$ emission from the BLR entirely and unattenuated. Having clarified this, in such a limit one would have that the intrinsic recombination radiation surface brightness is $(1 + \mathcal{R}_\textrm{max}(T))$ smaller than the observed one. As a consequence, all the densities and the masses discussed in the previous sections would need to be rescaled downwards by a factor $\sqrt{1 + \mathcal{R}_\textrm{max}(T)}$. For instance, for a distance $r = 50 \; \textrm{kpc}$, a temperature $T_\textrm{cold} = 5.6 \times 10^4 \; \textrm{K}$ and a volume filling factor $f_V = 10^{-3}$, the inferred density of the cold gas decreases from $n = 1.4 \; \textrm{cm}^{-3}$, in the recombination-dominated case, to $n = 0.23 \; \textrm{cm}^{-3}$, assuming maximum scattering. The neutral hydrogen column density and optical depth should also be rescaled, this time by a factor $1 + \mathcal{R}_\textrm{max}(T)$, due to the extra dependence of the neutral fraction of density (cf.\ Section \ref{subsec::H}).

\begin{figure}
\centering
\includegraphics[width=9cm]{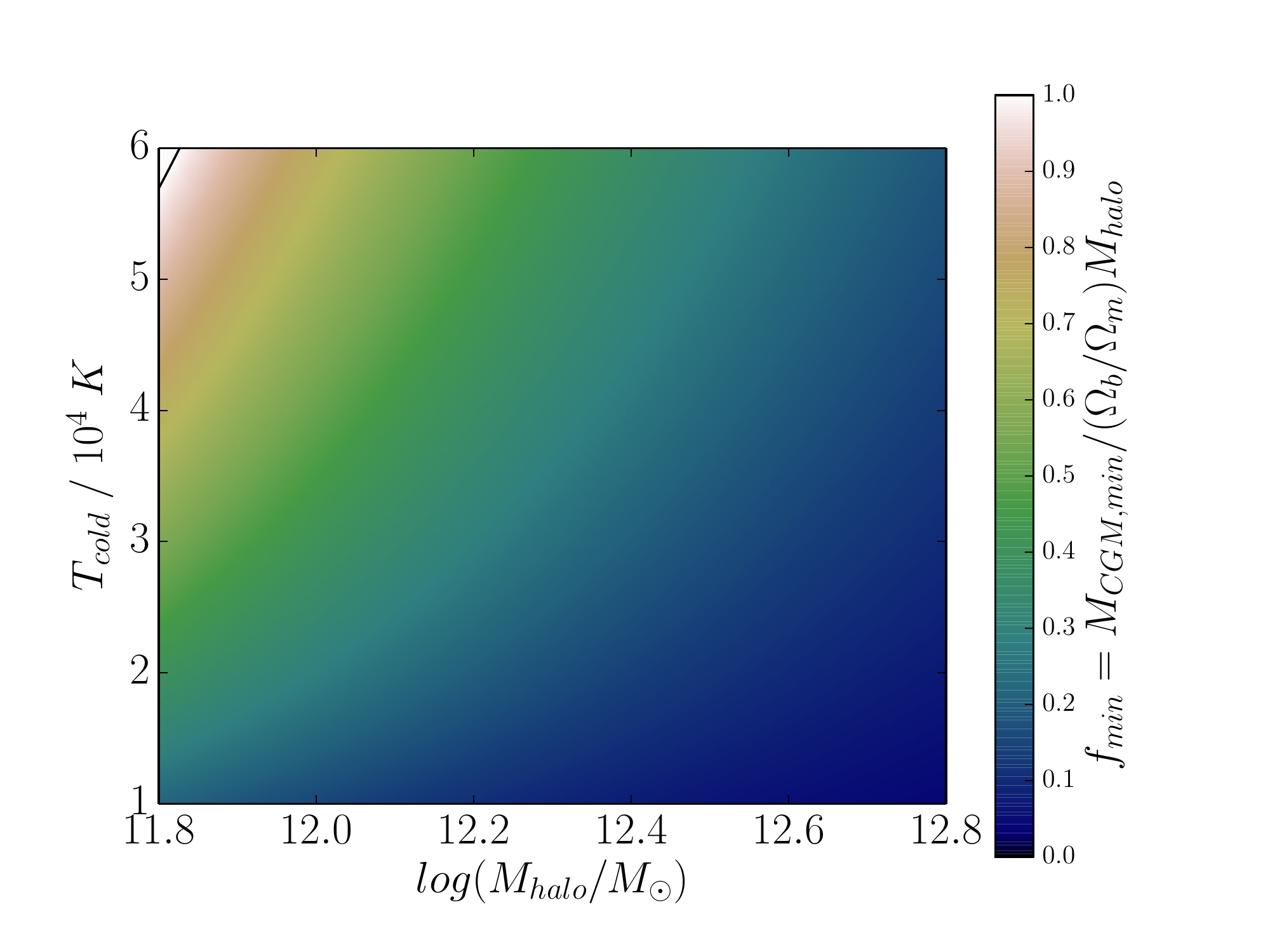}
\caption{Similar to Figure \ref{fig::minplot}, but assuming the \emph{maximum} possible contribution of scattering of BLR photons to the observed Ly$\alpha$ emission. Fiducial parameters ($M_\textrm{halo} = 10^{12} \; \textrm{M}_\odot$, $T_\textrm{cold} = 5.6 \times 10^4 \; \textrm{K}$; see Sections \ref{subsec::HaloMass} and \ref{subsec::Tcold}) correspond in this case to a \emph{minimum} CGM baryon fraction of 70\% of the cosmological value. Note that this model provides a very strict lower limit on the CGM baryon fraction, as it neglects any (possibly severe) attenuation of BLR photons while traversing the CGM (see the text for details).
\label{fig::minscattplot}}
\end{figure}

The minimum CGM baryon fraction assuming maximum possible contribution from scattering is given in Figure \ref{fig::minscattplot}, as a function of the assumed halo mass $M_\textrm{halo}$ and temperature of the cold gas $T_\textrm{cold}$. In this case, the fiducial values $M_\textrm{halo} \simeq 10^{12} \; \textrm{M}_\odot$ and $T_\textrm{cold} \simeq 5.6 \times 10^4 \; \textrm{K}$ (discussed in Sections \ref{subsec::HaloMass} and \ref{subsec::Tcold}, respectively) lead to a physically acceptable solution, with a \emph{minimum} CGM baryon fraction equal to 70\% of the cosmological value. This is still larger than the median baryon fraction predicted by many strong ejective feedback models (see references in Sections \ref{sec::Introduction} and \ref{subsec::minfbtot}). None the less, allowing for a few $10 \%$ of baryons in stars and ISM, this model is consistent with some moderate fraction of the baryons being expelled from haloes at higher redshift and not re-accreted yet. We recall that the room for expelled material gets lower if any of the two phases (cold or hot) dominates in mass over the other, as the inferred CGM baryon fraction is larger than the lower limit above if any of the two phases dominate over the other in mass (see Figure \ref{fig::totplot}). On the other hand, smaller CGM baryon fractions, similar to those predicted by strong ejective feedback models, can be obtained if just one of the two parameters $M_\textrm{halo}$ and $T_\textrm{cold}$ has a value much different from fiducial (rather than both of them, as in the recombination-dominated scenario; see Section \ref{subsec::minfbtot}).

We end this section by discussing the self-consistency and the plausibility of a scattering-dominated scenario. As mentioned above, the neutral hydrogen column density in this case needs to be rescaled, with respect to a recombination-dominated scenario, by a factor $1 + \mathcal{R}_\textrm{max}$. Equations \eqref{NHI} and \eqref{tau} therefore become
\begin{equation}
N_\textrm{HI} (r) = 2.7 \times 10^{13} \xi^{-1} \left( \dfrac{r}{10 \; \textrm{kpc}} \right)^{0.5} \; \textrm{cm}^{-2} \; ,
\end{equation}
and
\begin{equation}\label{tau_rescaled}
\tau_{\textrm{Ly}\alpha, \textrm{QSO}} (r) = 3.7 \times 10^{-2} \; \xi^{-1} \sigma_{300}^{-1} \left( \frac{r}{10 \; \textrm{kpc}} \right)^{0.5} \; ,
\end{equation}
respectively, where we omitted for clarity a negligible explicit dependence on temperature. Equation \eqref{tau_rescaled} in particular implies that a Ly$\alpha$ optically thin scenario is self-consistent out to a distance $r = 100 \; \textrm{kpc}$ for an intrinsic velocity dispersion as low as $\sigma = 35 \; \textrm{km} \; \textrm{s}^{-1}$ (assuming $\xi = 1$). 

We have therefore found that the recombination-dominated scenario and the scattering-dominated scenario are both formally self-consistent. As, however, they have distinctly different implications for the CGM baryon fraction (see Figures \ref{fig::minplot} and \ref{fig::minscattplot}), it would be desirable to find additional means to discriminate between these two options. 

One possible way to verify the viability of the scattering-dominated scenario could be to look for proximate absorption signatures in the QSO Ly$\alpha$ spectra. We have estimated that, in order to extract their luminosities from the BLR of the central QSOs, the MUSE Giant Ly$\alpha$ nebulae should offer to the central source, on average, a rest-frame equivalent width 
\begin{equation}\label{EW}
EW_0 = \frac{L_{\textrm{Ly$\alpha$}}}{L_{\lambda, \textrm{Ly}\alpha}^\textrm{QSO}} \left( \frac{\Omega}{4 \pi} \right)^{-1} =  0.3 \; \textrm{\r{A}} \; ,
\end{equation}
where we have used the median observed luminosity of the nebulae $L_{\textrm{Ly}\alpha} = 10^{44} \; \textrm{erg} \; \textrm{s}^{-1}$ (\citealt{Borisova+16}), a median QSO Ly$\alpha$ specific luminosity $L_{\lambda, \textrm{Ly}\alpha}^\textrm{QSO} = 3.3 \times 10^{44} \; \textrm{erg} \; \textrm{s}^{-1} \; \textrm{\r{A}}^{-1}$ (again assuming the typical QSO spectrum of \citealt{Lusso+15}) and we finally assumed that the nebulae surround their central QSOs completely, so that $\Omega = 4 \pi$. We have looked for associated absorption signatures in the spectra of the radio-quiet QSO of \cite{Borisova+16} within the wavelength range defined by the FWHM of the nebular spectra and we have found an average rest-frame equivalent width $EW_0 = 0.27 \; \textrm{\r{A}}$, fairly close to the expected value. We emphasize however that this kind of measurements should be taken with great caution, for a number of important reasons, including (i) the inherent difficulty of modelling absorption features on top of an emission line (here we simply modelled the `continuum' with a local linear interpolation), (ii) we cannot securely distinguish absorption features arising in the CGM from those associated with the BLR itself, the ISM of the host galaxy or even the IGM, so that the measurement should probably be interpreted as an upper limit and (iii) the down-the-barrel sightline offers a very limited and possibly biased view, which may not be representative of the average. We speculate however that studies of associated absorbers on larger samples and with a more detailed modelling of kinematics and radiative transfer could possibly help to distinguish between scenarios.

Finally, we want to emphasize that there is at least one example of a Giant Ly$\alpha$ nebula (the Slug nebula; \citealt{Cantalupo+14}) where a scattering-dominated scenario can be excluded at very high confidence. If applied to the Slug, a calculation similar to that in equation \eqref{EW} leads to a very different result than for the MUSE nebulae. This is due to (i) the relatively high luminosity of the Slug ($L_{\textrm{Ly}\alpha} = 2 \times 10^{44} \; \textrm{erg} \; \textrm{s}^{-1}$; \citealt{Cantalupo+14}) \footnote{Note that the illuminating QSO is instead very similar to those of the sample of \cite{Borisova+16}; see also \cite{FAB15}.} and most importantly (ii) to the fact that the Slug nebula is strongly asymmetric. Most of the luminosity arises from a region with size $l \simeq 80 \; \textrm{kpc}$ at a projected distance $R \simeq 80 \; \textrm{kpc}$ (\citealt{Cantalupo+14}). The solid angle subtended by the nebula as seen by the QSO can therefore be estimated as $\Omega \simeq 0.79 \; \textrm{srd}$, or $\Omega/4 \pi = 0.06$ (assuming that the true distance and the projected distance coincide) or even smaller (if the true distance is larger than the projected one, as argued for instance in \citealt{Cantalupo+19}, though note that we have little knowledge of the structure of this nebula along the line of sight). As a consequence, the \emph{rest-frame} equivalent width necessary to power the nebula with scattering is exceptionally high: $EW_0 = 9.6 \; \textrm{\r{A}}$, or possibly even larger that that. Note that, to be consistent with the optically thin assumption, this would require a velocity dispersion in excess of $1000 \; \textrm{km} \; \textrm{s}^{-1}$ (e.g.\ \citealt{Meiksin2009}). We can therefore conclude that scattering cannot be the dominant powering mechanism of the Slug nebula, as also found by the radiative transfer simulations of \cite{Cantalupo+14}. This is also in agreement with the recent detection, in the same object, of the non-resonant H$\alpha$ line, with a $\textrm{Ly}\alpha/\textrm{H}\alpha$ ratio consistent with recombination radiation (\citealt{Leibler+18}). Future H$\alpha$ observations, possible with the James Webb Space Telescope (JWST), will help to determine whether this is also the case for the MUSE Giant Ly$\alpha$ nebulae. Finally, polarization studies of Giant Ly$\alpha$ nebulae would also help constraining the contribution of scattering from a central source to the total observed emission (e.g.\ \citealt{Hayes+11}).

\subsection{Scattering within the nebula}\label{subsec::InternalScattering}

Some of the Ly$\alpha$ photons, coming either from recombinations or from the \emph{first} scattering of a BLR photon, can be scattered (or further scattered) within the nebula itself, thus affecting the surface brightness profile, while keeping the luminosity constant.\footnote{We neglect dust attenuation or any sort of net absorption, as they would only reinforce our conclusions; see Section \ref{subsec::caveats}.} A difference between the intrinsic and observed surface brightness profiles would imply a difference between the intrinsic density distribution of the cold gas and the one derived neglecting internal scattering (Section \ref{subsec::ncoldprofile}). Although the total luminosity is conserved in the process, it is not obvious that the inferred total mass (which is what we are interested in here) would remain unaffected as well. Here we show that this is indeed not the case, but that the required correction is small enough to have negligible impact to our conclusions.

Suppose that the `intrinsic' surface brightness distribution due to recombination is different from the observed one:
\begin{equation}\label{Sigmaoss}
\Sigma_\textrm{oss} (R) = \Sigma_{0, \textrm{oss}} \left( \frac{R}{R_0} \right)^{-\beta_\textrm{oss}}
\end{equation}
\begin{equation}\label{Sigmaint}
\Sigma_\textrm{int} (R) = \Sigma_{0, \textrm{intr}} \left( \frac{R}{R_0} \right)^{-\beta_\textrm{intr}}
\end{equation}
with $\Sigma_{0, \textrm{intr}} \neq \Sigma_{0, \textrm{oss}}$ and $\beta_\textrm{intr} \neq \beta_\textrm{oss}$. Then, using the notation of Section \ref{subsec::ncoldprofile}, the intrinsic density profile would be described by some intrinsic normalization and slope $(n_{0, \textrm{intr}}, \gamma_\textrm{intr})$ that are different from those that we derived from the observed profile, which we can call here $(n_{0, \textrm{oss}}, \gamma_\textrm{oss})$. 
We can integrate equations \eqref{Sigmaoss} and \eqref{Sigmaint} to obtain the total luminosity and the luminosity due to recombinations, respectively:
\begin{equation}\label{Ltot}
L_\textrm{tot} = \frac{8 \pi^2 \Sigma_{0, \textrm{oss}} R_0^2}{2 - \beta_\textrm{oss}} \left( x^{2 - \beta_\textrm{oss}} - 1 \right) \; ,
\end{equation}
\begin{equation}\label{Lrec}
L_\textrm{rec} = \frac{8 \pi^2 \Sigma_{0, \textrm{int}} R_0^2}{2 - \beta_\textrm{int}} \left( x^{2 - \beta_\textrm{int}} - 1 \right) \; ,
\end{equation}
where
\begin{equation}
x \equiv \frac{R_\textrm{max}}{R_0} \; .
\end{equation}
Note that in the integration above we use, as a lower limit, $R_0 = 10 \; \textrm{kpc}$, which is a reasonable estimate for the CGM/ISM transition, while $R_\textrm{max}$ is left free, but can be assumed to be of the order of the virial radius (equation \ref{Rvirdef}).

To derive the necessary corrections to the density profile, we then impose that the total luminosity be
\begin{equation}\label{Lbalance}
L_\textrm{tot} = (1 + f_\textrm{BLR}) L_\textrm{rec} \; , 
\end{equation}
where $f_\textrm{BLR}$ is the \emph{global} average of the ratio defined in equation \eqref{jratio} and will therefore vary in the range $0 < f_\textrm{BLR} < \mathcal{R}_\textrm{max}$ (see equation \ref{Rdef}). Combining equations \eqref{Ltot}, \eqref{Lrec} and \eqref{Lbalance} with equations \eqref{n0}, we find
\begin{equation}\label{nratio}
\frac{n_{0, \textrm{intr}}}{n_{0, \textrm{oss}}} = \frac{1}{\sqrt{1+f_\textrm{BLR}}} \left ( \frac{\chi(1+\beta_\textrm{oss})}{\chi(1+\beta_\textrm{intr})} \;  \frac{2 - \beta_\textrm{intr}}{2 - \beta_\textrm{oss}} \; \frac{x^{2-\beta_\textrm{oss}}-1}{x^{2-\beta_\textrm{intr}}-1} \right)^{1/2} \; .
\end{equation}
The total cold gas mass is then found integrating over radius (as in Section \ref{subsec::fbcold}), thence
\begin{equation}\label{Mratio}
\frac{M_\textrm{intr}}{M_\textrm{oss}} = \frac{3 - \gamma_\textrm{oss}}{3 - \gamma_\textrm{intr}} \; \frac{n_{0, \textrm{intr}}}{n_{0, \textrm{oss}}} x^{\gamma_\textrm{oss}-\gamma_\textrm{intr}} \; ,
\end{equation}
where the $\gamma$'s are as in equation \eqref{n0}.

\begin{figure}
\centering
\includegraphics[width=9cm]{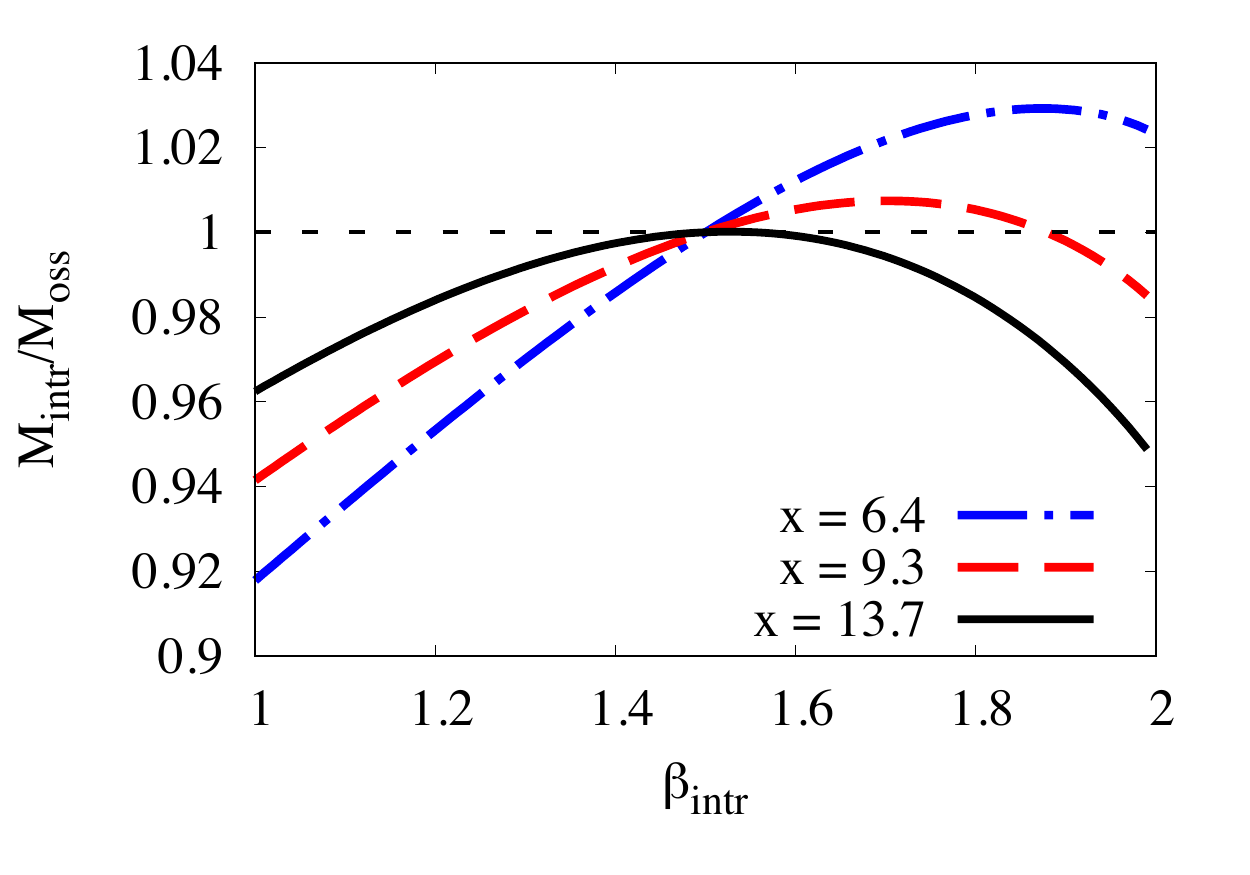}
\caption{Correction factor to the estimated CGM mass, due to internal scattering within the nebula. Results are given as a function of the intrinsic slope $\beta_\textrm{intr}$ of the pure recombination Ly$\alpha$ surface brightness profile, and for various values of the geometrical parameter $x$ describing the radial extent of the nebula (see the text for details). Corrections amount to unity (by construction) for $\beta_\textrm{intr} = \beta_\textrm{oss} = 1.5$. The bias due to internal scattering is in all cases smaller than 10\% and has therefore negligible impact on our conclusions.\label{fig::Mcorrection}}
\end{figure}

The final correction factor on the cold gas mass, which then propagates to a correction to the total (cold plus hot) inferred mass of the CGM, is readily obtained by combining equations \eqref{nratio} and \eqref{Mratio}. The resulting expression (which we do not write explicitly as it is as long as trivial) contains a factor $(1 + f_\textrm{BLR})^{-1/2} \leq (1 + \mathcal{R}_\textrm{max})^{-1/2}$, which is just the same that was discussed already in Section \ref{subsec::BLR}. The remaining part of the expression gives any additional bias in the determination of the CGM mass due to internal scattering within the nebula. This is visualized in Figure \ref{fig::Mcorrection}, as a function of the intrinsic slope $\beta_\textrm{intr}$, for various values of the radial range parameter $x$, chosen to comprise the range of virial radii plausible for our haloes. Though the internal scattering of recombination radiation would only act in making the profile shallower, we consider here both the options $\beta_\textrm{intr} < \beta_\textrm{oss}$ and $\beta_\textrm{intr} > \beta_\textrm{oss}$, as the first scattering from the BLR can instead result in a centrally concentrated contribution in the high optical depth regime (large $\tau_{\textrm{Ly}\alpha, \textrm{QSO}}$, which however corresponds to small $\mathcal{R}$ and $f_\textrm{BLR}$, as discussed in Section \ref{subsec::BLR}). It is clear from the figure that the correction is of the order of a few per cent at most and therefore has no significant impact on our results.

\subsection{Scattering off a neutral screen}\label{subsec::NeutralScreen}

An additional redistribution effect, which can be partly treated with the same formalism of Section \ref{subsec::InternalScattering}, would also arise if some of the Ly$\alpha$ photons originated in the CGM and travelling \emph{away} from the observer are scattered back into the line of sight after interacting with a layer or region of material with a relatively high neutral fraction and neutral hydrogen column density. This can happen, for instance, if the central QSO is obscured along some directions away from the line of sight, so that some regions of the CGM are less ionized than others and offer higher neutral hydrogen column density, as suggested, for instance, by absorption studies of the CGM (\citealt{QPQ6}) around QSOs of similar redshift as (although about 1 dex less luminous than) those of \cite{Borisova+16}.

This process can change the shape of the surface brightness profile -- an effect that we addressed already in Section \ref{subsec::InternalScattering} -- but, in addition to that, it could also introduce a global normalization bias, as light is being added along some lines of sight. At the same time, however, this potential normalization enhancement would be counteracted by the fact that the whole `far' side of the nebula would be largely obscured by the intervening screen and therefore not contribute to the observed Ly$\alpha$ flux. Insight suggests that the latter effect should in general be stronger than the former, so that the combination of the two is unlikely to bias the total flux \emph{high} and therefore to weaken our conclusions. The idea can be intuitively illustrated as follows.

Suppose that the QSO is able to ionize only a fraction $f_\textrm{ion} < 1$ of the entire solid angle. We include in this both the volume ionized by the QSO directly and the volume ionized indirectly by ionizing photons produced in recombination events (e.g.\ \citealt{OsterbrockFerland}). Then only a fraction $0.5 f_\textrm{ion}$ of the CGM volume would be bright in recombination radiation (or capable to directly see and scatter photons from the BLR) \emph{and} also be directly visible by the observer. At the same time, some fraction $f_\textrm{screen}$ of the photons produced in this region would intercept the screen, where they can be scattered back (and possibly reach the observer) with some probability $f_\textrm{back}$, or find their way through the screen with a probability $1 - f_\textrm{back}$. Also, with a probability $f_\textrm{screen}(1- f_\textrm{back})$, photons from the `far' side of the CGM would intercept and traverse the screen and also contribute to the observed flux. The total observed Ly$\alpha$ flux would be different from that of an ideal case with perfect illumination, by a factor
\begin{equation}
\left. \begin{array}{ll}
\dfrac{F_\textrm{oss}}{F_\textrm{ideal}} & = \dfrac{1}{2} f_\textrm{ion} (1 + f_\textrm{screen} f_\textrm{back}) + \dfrac{1}{2} f_\textrm{ion} f_\textrm{screen} (1 - f_\textrm{back})  \\
& \\
& = \dfrac{1}{2} f_\textrm{ion} ( 1 + f_\textrm{screen} ) \leq 1 \; .
\end{array} \right. 
\end{equation}
Although highly idealized, the inequality above suggests that, if anything, the presence of a neutral screen is more likely to introduce a bias in the direction of strengthening our results, rather than viceversa. We plan to demonstrate this point more thoroughly with radiative transfer simulations in future work.

\section{Discussion}\label{sec::Discussion}

We discuss here the following interrelated questions: (i) What are the typical size and the number of the cold gas structures that produce the emission in MUSE Giant Ly$\alpha$ nebulae? (ii) What is the origin of the cold gas? (iii) How well justified is our assumption of pressure equilibrium with the hot gas? When relevant, we also try to connect our discussion to resolution or other numerical effects that may be important for the interpretation of cosmological hydrodynamical simulations of the CGM.

\subsection{The typical size of the cold gas structures}
So far, we have not addressed the question of what is the typical size and the number of the cold gas structures in the MUSE Giant Ly$\alpha$ nebulae. This is because the main parameter in our calculations is the volume filling factor $f_V$ and any given volume filling factor can be obtained either with a large number of small clouds or with a smaller number of larger ones. One can however attempt to break the degeneracy, at least qualitatively, by considering the \emph{area covering factor} $f_C$, which is more easily accessible to observations than $f_V$ and is given by
\begin{equation}\label{fC}
f_C = \int \frac{f_V}{R_c} dl \; ,
\end{equation}
where the integral is along the line of sight and $R_c$ is the typical size of the individual cold gas structures along the same direction. Assuming that the intrinsic orientation of the structures is independent on our line of sight, we can take $R_c$ as the typical physical scale of the emitting structures, averaged over all directions; we stress that $R_c$ does not necessarily represent the radius of spherical clouds, as we did not make any assumptions on the geometry and not even on the topology of the emitting structures.

As an order-of-magnitude estimate, equation \eqref{fC} provides
\begin{equation}\label{Rcestimate}
R_c \simeq f_V r_\textrm{vir} f_C^{-1} \lesssim 74 \left( \frac{f_V}{10^{-3}} \right) \; \textrm{pc} \; ,
\end{equation}
where we assumed that the typical length of the line of sight through the CGM is of the order of the virial radius (equation \ref{Rvirdef}) and we also assumed $f_C \gtrsim 1$. While we do not know what the true covering factor of the cold CGM is (due to the seeing-limited spatial resolution of MUSE), the continuity and smoothness of the emission suggest that it is not much smaller than unity. {\cite{FAB15} have shown with synthetic observations that, even accounting for finite resolution, the observed morphology of extended Ly$\alpha$ emission requires an intrinsic covering factor $f_C > 0.5$. Large covering factors are also supported by absorption studies, which are more sparse but have virtually infinite angular resolution (e.g.\ \citealt{QPQ6}). Note that $f_C > 1$ is possible if multiple structures overlap in projection along the line of sight, which explains why equation \eqref{Rcestimate} is best interpreted as an upper limit. We stress that the one above is just an order-of-magnitude estimate and it is only useful as a guidance to discuss which scenarios for the nature and origin of the cold gas should be considered more plausible than others.

\subsection{Most of the cold gas is not in gravitationally bound substructures}\label{subsec::selfgravity}
Using the first equality in \eqref{Rcestimate} and the definition of volume filling factor, one can also estimate (again by order of magnitude) what number of cold structures with size $R_c$ can be found within the virial radius:
\begin{equation}
N \simeq f_V^{-2} f_C^{3} \gtrsim 10^4 \; ,
\end{equation}
assuming $f_V \lesssim 10^{-2}$ and $f_C \gtrsim 1$. This simple estimate disfavours a scenario in which the cold gas structures are individually associated with dark matter sub-haloes (and therefore suggests that the Ly$\alpha$ emission cannot originate, for instance, from ISM bound to undetected satellite galaxies), as the latter are only a few hundreds at most (e.g.\ \citealt{Moore+99}; \citealt{RodriguezPuebla+16}). We recall that larger values of the filling factor $f_V$ would require an implausibly large mass of cold gas to explain the observed surface brightness (see Section \ref{subsec::fbcold}); note also that this conclusion is independent on assumptions about pressure confinement.

The size estimate \eqref{Rcestimate} also helps determining whether the cold gas structures are collapsed or on the verge of collapsing under their own self-gravity. This is an intriguing option, as it could imply that some of the Ly$\alpha$ emitting structures could be related to precursors of globular clusters. To this aim, we should compare $R_c$ to the Jeans length
\begin{equation}\label{lambdaJ}
\lambda_J = \sqrt{\frac{\pi c_s^2}{G \rho}} = 16 \; T_4^{0.02} f_V^{1/4} \; \textrm{kpc} \; ,
\end{equation}
where again $T_4 \equiv T / 10^4 \; \textrm{K}$, $c_s$ is the sound speed of an ionized plasma at temperature $T$ and we used equation \eqref{densprofile}, evaluated at a reference radius of 50 kpc. The negligible dependence on temperature originates from a combination of the temperature dependencies of the sound speed and the effective Ly$\alpha$ recombination coefficient. Combining equations \eqref{Rcestimate} and \eqref{lambdaJ} gives the ratio
\begin{equation}
r_J \equiv \frac{R_c}{\lambda_J} \lesssim 4.6 \; T_4^{-0.02} f_V^{3/4} \; .
\end{equation}
We see that the condition $r_J < 1$ for stability is safely met for any filling factor of interest. We conclude that the cold gas structures are most likely stable against gravitational collapse.
In the calculation above, we used our density estimate based on a pure recombination scenario (Section \ref{sec::coldCGM}), but note that our conclusion would only be reinforced if scattering is non-negligible, as the estimated densities decrease in that case (Section \ref{subsec::BLR}), further enhancing stability against gravitational collapse.

\subsection{Can the cold gas be overpressured?}\label{subsec::overP}

As the gravitational pull (by either dark matter sub-haloes or the gas itself) can be neglected (see Section \ref{subsec::selfgravity}), we are confident that the cold gas structures are most likely confined by the pressure of the surrounding medium, which justifies the main assumption of our Section \ref{sec::Hot}. We also remind that such external pressure can include non-thermal contributions, such as those due to large-scale turbulence, and that the precise amount of thermal and non-thermal pressure has little impact on our results. With this specification in mind, any residual pressure imbalance should be smeared out roughly on one sound crossing time-scale
\begin{equation}\label{tsound}
t_\textrm{sound} = \frac{R_c}{c_s} = 2.0 \; \left( \frac{T_\textrm{cold}}{5.6 \times 10^4 \; \textrm{K}} \right)^{-1/2} \left( \frac{R_c}{74 \; \textrm{pc}} \right) \; \textrm{Myr} \; ,
\end{equation}
where we scaled the result to the reference temperature and size estimated by means of equations \eqref{Teq} and \eqref{Rcestimate}, respectively.

One possibility to explain a large Ly$\alpha$ surface brightness without resorting to large baryon fractions could therefore be that the majority of the cold gas structures were either created, or they acquired their current pressure, less than $\sim 2 \; \textrm{Myr}$ ago. There is one simple way to achieve this. If the QSO lifetime is significantly smaller than $\sim 2 \; \textrm{Myr}$, the cold gas that we see bright in Ly$\alpha$ should be modelled as having been suddenly heated to the equilibrium temperature (see Section \ref{subsec::Tcold}) and may still find itself temporarily overpressured with respect to the ambient medium, by an amount $T_\textrm{cold}/T_\textrm{in}$, where $T_\textrm{in}$ is its temperature before being illuminated by the QSO. Quantitatively, this \emph{sudden photo-heating} scenario is equivalent to rescaling the factor $\varepsilon$ in equation \eqref{epsilon} by $T_\textrm{in}/T_\textrm{cold}$. As a result, equation \eqref{mumin} for the \emph{minimum} CGM baryon fraction reads in this case
\begin{equation}
f_\textrm{min} = 2 f_1 \varepsilon_1^{1/2} T_{\textrm{in}, 4}^{1/2} T_4^{t/2} M_{12}^{-(1+\gamma)/3} \; ,
\end{equation}
with $T_{\textrm{in}, 4} \equiv T_\textrm{in} / 10^4 \; \textrm{K}$ and the other quantities as in Section \ref{subsec::minfbtot}. The minimum CGM baryon fraction for this model is shown in Figure \ref{fig::suddenphotoheating} assuming an \emph{initial} cold gas temperature $T_\textrm{in} = 10^4 \; \textrm{K}$ and recombination-dominated emission. Comparison to Figure \ref{fig::minplot} shows that sudden photo-heating facilitates the achievement of a physically acceptable solution, without requiring uncomfortable assumptions on the QSO spectrum or CGM metallicity (Section \ref{subsec::Tcold}), although relatively large halo masses $M_\textrm{halo} \gtrsim 10^{12.4} \; \textrm{M}_\odot$ are still required. 

\begin{figure}
\centering
\includegraphics[width=9cm]{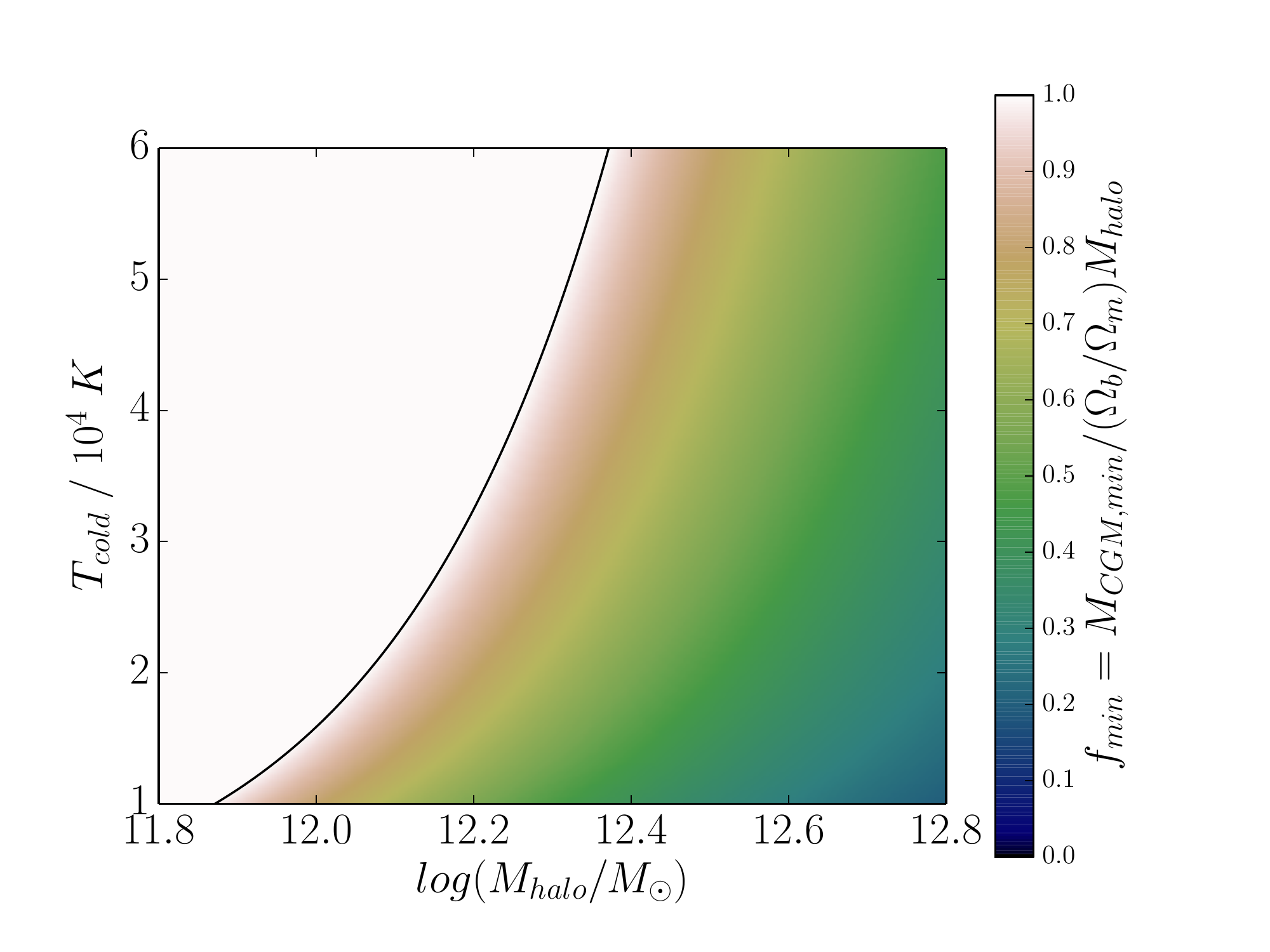}
\caption{Similar to Figure \ref{fig::minplot}, but assuming that the central QSO has been `on' for less than $\sim 2 \; \textrm{Myr}$, so that the suddenly photo-heated cold gas did not have the time yet to respond to the pressure imbalance and expand in the surrounding medium. \label{fig::suddenphotoheating}}
\end{figure}

Short-lived overpressures could also originate, for instance, if some of the cold gas is ballistically ejected from the ISM of a galaxy, without significantly interacting with the surrounding medium before reaching a considerable distance within the CGM. A discussion of the conditions under which this can happen would go beyond the scope of this paper. We note, however, that schemes corresponding to this description are sometimes implemented in some recipes for feedback in cosmological hydrodynamical simulations (e.g.\ \citealt{Vogelsberger+13} and references therein). We therefore expect some amount of overpressured gas to be found in such simulations, but we warn that it can be, at least to some extent, an unphysical consequence of the adopted hydrodynamic/feedback scheme and should therefore be considered with caution.

One should finally consider the option that the cold gas structures are \emph{internally} turbulent. Super-sonic turbulence would induce a lognormal density distribution \emph{within} the cold gas structures (e.g.\ \citealt{Molina+12} and references therein), leading to an additional \emph{internal} clumping factor (see Section \ref{subsubsec::fV}), which we did not account for in our calculations, though there are some recent observational indications pointing in this direction (\citealt{Cantalupo+19}). Significant internal clumpiness would have an impact on the estimated CGM baryon fraction of the cold gas. For instance, following \cite{Nolan+15}, internal supersonic turbulence with a Mach number $M = 2.5$ would induce in the cold gas a lognormal density distribution with dispersion $\sigma_\textrm{ln} = 0.7$. This would imply an internal clumping factor $C_\textrm{int}  = e^{\sigma_\textrm{ln}^2} = 2$ and drive our mass estimates of the cold gas down by a factor of 2. Two potential problems are worth being pointed out concerning this scenario. First, internal turbulence would imply a total (thermal plus turbulent) \emph{internal} pressure that is about $(1 + \Gamma M^2)$ larger than the thermal component alone (with $\Gamma = 5/3$ for a monoatomic gas). This would require an equal enhancement of the \emph{external} confining pressure and imply an increase of the estimated mass of the hot gas (e.g.\ by a factor of 11 for $M = 2.5$), more than compensating the decrease in the estimated mass of the cold gas. The last conclusion does not hold, of course, if the phenomenon is short-lived and set to be smeared out on a turbulent crossing time. Note that for supersonic turbulence the turbulent crossing time would be even smaller than the sound crossing time in equation \eqref{tsound}, by a factor of $\sim M$, so that the emitting structures should have become turbulent, in this scenario, roughly 1 Myr ago. Also note, in contrast to the `sudden photo-heating' scenario discussed above, that it is not clear why supersonic turbulence should have been set on so recently. Second, supersonic turbulence would quickly heat the cold gas, through repeated shocks, to high temperatures, where the Ly$\alpha$ emissivity drops, further lowering the survivability of this situation, by an amount that we did not try to quantify.
We cannot exclude that a large internal clumpiness can be due to processes other than turbulence. We are not aware, however, of plausible alternatives. We recall that self-gravity (which would in principle be an appealing candidate) can be neglected here, as we have argued already in Section \ref{subsec::selfgravity}.

\subsection{Progressive fragmentation and incomplete pressurization}

Although persistent overpressures are challenging to explain, an underpressured cold CGM can instead be achieved under somewhat easier circumstances, which we briefly discuss below. 

Cold gas accreting onto a halo from IGM filaments can find itself, at the moment of entering the halo, with a lower pressure with respect to the hot gas already present there (see e.g.\ the phase diagrams in \citealt{Nelson+16}). The hot gas will then compress the incoming gas to higher densities, but the process is not necessarily well approximated as being instantaneous. It is therefore important to establish whether the incoming gas has the time to be pressurized before penetrating deep into the halo. One way to do this is to compare the sound-crossing time of a filament
\begin{equation}\label{tsoundfil}
t_\textrm{sound, fil} = \frac{R_\textrm{fil}}{c_s} = 640 \; T_4^{-1/2} \left( \frac{R_\textrm{fil}}{10 \; \textrm{kpc}} \right) \; \textrm{Myr}
\end{equation}
and its halo crossing time
\begin{equation}
t_\textrm{cross, fil} = \frac{r_\textrm{vir}}{V_\textrm{fil}} = 980 \left( \frac{r_\textrm{vir}}{100 \; \textrm{kpc}} \right) \; \left( \frac{V_\textrm{fil}}{100 \; \textrm{km} \; \textrm{s}^{-1}} \right)^{-1} \textrm{Myr} \; .
\end{equation}
It is readily seen that sufficiently thick or fast filaments could in principle be able to travel significant distances within the CGM before being completely pressurized by the hot gas. Part of this gas would therefore be underpressured and emit less Ly$\alpha$ than otherwise.

It is possible and even likely, however, that pressurization proceeds much faster then the simple estimate above. Unless perfect symmetry conditions are met, in fact, an underpressured structure would be progressively deformed, during compression, and possibly fragment in smaller units. Fragments (or highly deformed portions of the original structure) will have a smaller sound-crossing time and therefore be prone to even faster compression and further fragmentation. The process has a run-away nature and results in very rapid progressive fragmentation in structures with smaller and smaller size, until complete pressurization is reached, or some other process (e.g.\ thermal conduction) takes over, either halting further fragmentation or destroying the fragments altogether.\footnote{This process should not be confused with the interaction between the hot gas and an inflowing filament \emph{assumed} to be in pressure equilibrium since the beginning and which can then be disrupted by the Kelvin-Helmholtz instability; this can result in moderate local deviations from pressure equilibrium (e.g.\ \citealt{Mandelker+16}; \citealt{Padnos+18}; Vossberg et al.\ in prep.)}

It is also possible that some of the cold gas originates from the condensation of the hot gas itself. Progressive fragmentation is indeed usually described in the context of thermal instability, in which radiative cooling is the means to create local underpressures and initiate fragmentation. The final size of the cold gas structures in this case is roughly given by the \emph{acoustic length}, for which the cooling time equals the sound crossing time (e.g.\ \citealt{InoueOmukai2015}). \cite{McCourt2018} recently described the phenomenon under the name of \emph{shattering} and clarified that fragmentation would stop at the \emph{cooling length} (the minimum acoustic length achieved during the cooling process), unless this is smaller than the Field length, below which thermal conduction would lead the cold gas to evaporate. We also notice that, if the radial profiles of heating and cooling meet certain conditions (specifically, if the heating rate declines with radius more steeply than the cooling rate), some cold gas can condensate in a shell of finite radius, possibly meeting the requirements for the onset of the Rayleigh-Taylor instability and eventually lead to the creation of filamentary structures within the CGM.\footnote{We thank A.\ Negri for bringing this option to our attention.} 

Crucially, high resolution is required for the processes mentioned above to be captured in a hydrodynamical simulation. The cooling length, in particular, can be as small as a fraction of a pc (\citealt{McCourt2018}). It is therefore possible that underpressured gas may be present in some cosmological simulations, as a consequence of non-negligible nominal compression time-scales (equation \ref{tsoundfil}), but it is also possible that some of these features may partially be artefacts, due to the numerical resolution being insufficient to follow the shattering process and the resultant accelerated compression. Note that this last consideration should mostly apply to grid codes. SPH codes could suffer from different issues, including on one side spurious clumpiness (e.g.\ \citealt{Torrey+12}) and on the other side artificial underpressures associated to the non-trivial nature of density evaluations in SPH codes (see e.g.\ Appendix D in \citealt{Oppenheimer+18}).

We conclude by emphasizing that a systematically underpressured cold CGM, if real, would imply that the hot halo is even more massive than what inferred based on pressure balance, and would therefore go in the direction of strengthening our results.

\section{Summary and conclusions}\label{sec::Summary}

Galaxy formation models rely on mechanisms to keep most of the baryons outside of galaxies, but different scenarios make different predictions about how far away from galaxies the gas is expected to be found. Measuring what fraction of baryons is retained within the virial radius of haloes -- especially at high redshift, when most of the relevant processes are expected to take place -- is therefore crucial to inform and constrain theoretical models and especially different flavours of feedback.

In this work, we have tried to determine the total baryonic mass stored in the CGM of relatively massive ($M_\textrm{halo} \sim 10^{12} \; \textrm{M}_\odot$) haloes at $z\sim3$, based on the observed surface brightness profile of Giant Ly$\alpha$ nebulae discovered with MUSE around luminous QSOs at this redshift (\citealt{Borisova+16}). To this aim, we modelled the CGM as a multiphase medium, where a hot (collisionally ionized) plasma confines, due to its pressure, a colder (photo-ionized) phase into a small fraction of the volume and to large densities, with correspondingly large Ly$\alpha$ emissivity. 

We have found that a CGM dominated in mass by any of the two phases would fall short in reproducing the observed surface brightness, either because the cold emitting gas is not enough, or because it has too low density (and therefore too low emissivity) due to insufficient compression from the ambient medium. On the other hand, by assuming that the two phases provide similar contributions to the total mass (i.e.\ by maximizing the theoretical radiative output of the model), one can put a lower limit to the total mass of the CGM required to explain the observed surface brightness. Our CGM mass estimates are lower limits also because we have neglected a number of effects (dust extinction, IGM attenuation, partial illumination due to a narrow QSO ionization cone), which would diminish the predicted Ly$\alpha$ luminosity and therefore increase the amount of gas needed to explain the observations.

We have thereby found that a large total (hot plus cold) baryonic mass of the CGM -- close to the total amount of baryons nominally associated with these haloes -- is required to give account for a Ly$\alpha$ surface brightness as high as observed. The physical constraint that the baryon fraction be equal to or smaller than the cosmological value requires \emph{at least one} of the following four conditions to be met:

\begin{enumerate}
\item{The mass of QSO-hosting dark matter haloes at $z \sim 3$ is larger than the current best estimates based on clustering ($M_\textrm{halo} \sim 10^{12} \; \textrm{M}_\odot$; Section \ref{subsec::HaloMass});}
\item{The equilibrium temperature of the photo-ionized gas is significantly smaller than what expected in a `standard' QSO ionizing spectrum ($T_\textrm{eq} \sim 5 \times 10^4 \; \textrm{K}$; Section \ref{subsec::Tcold});}
\item{The intrinsic radial velocity dispersion of the CGM is large enough that the photo-ionized gas is optically thin to Ly$\alpha$ scattering (Section \ref{subsec::BLR});}
\item{Most of the cold gas is in a short-lived overpressured phase, which we observe before it had the time to expand in the surrounding medium (Section \ref{subsec::overP}).}
\end{enumerate}
All these options can potentially be constrained by independent observations. Particular help would come from further constraints on clustering and spectral properties of high-$z$ luminous QSOs, as well as from follow-up studies on CGM non-resonant emission lines and/or polarimetry of the Ly$\alpha$ emission. The last possibility (short-lived overpressures) is the most difficult to test, but we estimated that it could be fulfilled if the majority of bright QSOs at $z = 3$ have a life-time shorter than $\sim 2 \; \textrm{Myr}$.

\emph{At least two} of the conditions above must be met, if the observed surface brightness is to be reconciled with the relatively small CGM baryon fractions ($30-40\%$ of the cosmic average) predicted at the relevant halo mass and redshift by the extremely strong ejective feedback sometimes adopted in galaxy formation models. More comparisons with models and simulations would be desirable to derive more robust conclusions. Regarding the comparison of QSO Giant Ly$\alpha$ nebulae with simulations, we stress in particular the importance of explicitly including photo-electric heating due to photo-ionization of HeII, as this significantly impacts the temperature, recombination coefficient and ultimately the emissivity of the gas.

In conclusion, our results seem to indicate a preference for a relatively high baryon fraction (close to cosmological) in the CGM of massive haloes at $z\sim3$, although more observations are needed to alleviate the existing uncertainties. If confirmed, our findings could help adding constraints on fundamental parameters of galaxy formation models, such as the efficiency with which SN feedback can expel baryons from galactic haloes in the early Universe.

\section*{Acknowledgements}
We thank Arif Babul, Andrea Ferrara, Max Gronke, Andrea Negri and Benny Trakhtenbrot for useful discussion, Raffaella Anna Marino for help and support and Peter Mitchell for sharing some of his analysis of the EAGLE simulations. This work was supported by the Swiss National Science Foundation grant PP00P2\_163824.

\bibliographystyle{mn2e}
\bibliography{mybib}{}

\label{lastpage}
\end{document}